\documentclass[]{aa}
\usepackage[varg]{txfonts}
\usepackage{twoopt} 

\usepackage[vlined]{algorithm2e} 
\SetAlgoCaptionSeparator{.}

\usepackage[%
  breaklinks = true,
  colorlinks = true,
  urlcolor   = blue,
  citecolor  = blue,
  linkcolor  = blue,
]{hyperref}

\bibpunct{(}{)}{;}{a}{}{,} 

\graphicspath{{figures_paper/}} 

\makeatletter
 \newcommandtwoopt{\citeads}[3][][]{%
   \nonstopmode
   \href{http://adsabs.harvard.edu/abs/#3}%
        {\def\hyper@linkstart##1##2{}%
         \let\hyper@linkend\@empty\citealp[#1][#2]{#3}}
   \biblink{#3}{\href{http://adsabs.harvard.edu/abs/#3}{ADS}}%
   \errorstopmode}            
 \newcommandtwoopt{\citepads}[3][][]{%
   \nonstopmode
   \href{http://adsabs.harvard.edu/abs/#3}%
        {\def\hyper@linkstart##1##2{}%
         \let\hyper@linkend\@empty\citep[#1][#2]{#3}}
   \biblink{#3}{\href{http://adsabs.harvard.edu/abs/#3}{ADS}}
   \errorstopmode}            
 \newcommandtwoopt{\citetads}[3][][]{%
   \nonstopmode
   \href{http://adsabs.harvard.edu/abs/#3}
        {\def\hyper@linkstart##1##2{}%
         \let\hyper@linkend\@empty\citet[#1][#2]{#3}}
   \biblink{#3}{\href{http://adsabs.harvard.edu/abs/#3}{ADS}}%
   \errorstopmode}            
 \newcommandtwoopt{\citeyearads}[3][][]{%
   \nonstopmode
   \href{http://adsabs.harvard.edu/abs/#3}%
        {\def\hyper@linkstart##1##2{}%
         \let\hyper@linkend\@empty\citeyear[#1][#2]{#3}}
   \biblink{#3}{\href{http://adsabs.harvard.edu/abs/#3}{ADS}}%
   \errorstopmode}            
\makeatother
\makeatletter
\newcommand{\bibnote}[2]{\@namedef{#1note}{#2}}
\newcommand{\biblink}[2]{\@namedef{#1link}{#2}}
\makeatother

\newcommand{\CaIIHK}{\ion{Ca}{II}\,H\&K}
\newcommand{\MgIIHK}{\ion{Mg}{II}\,h\&k}

\newcommand{\CaIR}{\CaII~8542 \AA}

\newcommand{\MgIIh}{\ion{Mg}{II}\,h}

\newcommand{\CaIIK}{\ion{Ca}{II}\,K}
\newcommand{\MgIIk}{\ion{Mg}{II}\,k}

\newcommand{\CaII}{\ion{Ca}{II}}
\newcommand{\MgII}{\ion{Mg}{II}}
\newcommand{\HI}{\ion{H}{I}}

\newcommand{\lyalpha}{Ly$\alpha$}
\newcommand{\halpha}{H$\alpha$}

\begin{document}

\title{Three-dimensional modeling of chromospheric spectral lines in a simulated active region}

\author{
  Johan P. Bj{\o}rgen\inst{\ref{ISP}},
  Jorrit Leenaarts\inst{\ref{ISP}},
  Matthias Rempel\inst{\ref{HAO}},
  Mark C. M. Cheung\inst{\ref{LMSAL}},
  Sanja Danilovic\inst{\ref{ISP}}, 
  Jaime de~la Cruz Rodr\'{\i}guez\inst{\ref{ISP}}, and
  Andrii V. Sukhorukov \inst{\ref{IAC},\ref{MAO}}
}

\institute{
  Institute for Solar Physics,
  Department of Astronomy, Stockholm University, AlbaNova University Centre,
  SE-106~91 Stockholm, Sweden
  \label{ISP}
  \and
  High Altitude Observatory,
  NCAR, P.O. Box 3000, Boulder, Colorado 80307, USA
  \label{HAO}
  \and
  Lockheed Martin Solar and Astrophysics Laboratory, 3251 Hanover Street Bldg. 252, Palo Alto, CA 94304, USA
  \label{LMSAL}
  \and
  Instituto de Astrof\'{\i}sica de Canarias, E-38205 La Laguna, Tenerife, Spain
  \label{IAC}
  \and 
  Main Astronomical Observatory, National Academy of Sciences of Ukraine,
  27 Akademika Zabolotnoho Str., 03680 Kyiv, Ukraine
  \label{MAO}
  }

\offprints{J. P. Bj{\o}rgen, \email{johan.bjorgen@astro.su.se}}

\date{Received; Accepted }

\abstract{
{Because of the complex physics that governs the formation of chromospheric lines, interpretation of solar chromospheric observations is difficult. The origin and characteristics of many chromospheric features are, because of this, unresolved.}
}{
  {We focus here on studying two prominent features: long fibrils and flare ribbons. To model them, we use a 3D MHD simulation of an active region which self-consistently reproduces both of them.}
}{
  We model {the \halpha, \MgIIk, \CaIIK, and Ca \textsc{ii} 8542 \AA\ } lines using the 3D non-LTE radiative transfer code Multi3D.
  To obtain non-LTE electron densities, we solve the statistical equilibrium equations for hydrogen simultaneously with the 
   charge conservation equation. We treat the \CaIIK\ and \MgIIk\ lines with partially coherent scattering.
}{
	This simulation reproduces long fibrils that span between the opposite-polarity sunspots {and go up to 4 Mm in height. They can be traced in all lines due to density corrugation. Opposite to previous studies, \halpha, \MgIIHK, and \CaIIHK, are formed at similar height in this model. Although, some of the high fibrils are also visible in  the Ca \textsc{ii} 8542 \AA\ line, this line tend to sample  loops and shocks lower in the chromosphere.} Magnetic field lines are aligned with the \halpha\  fibrils, but the latter holds to a lesser extent for the {Ca \textsc{ii} 8542 \AA\ line}. 

The simulation shows structures in the \halpha\ line core that look like flare ribbons. The emission in the ribbons is caused by a dense chromosphere and a transition region at high column mass. The ribbons are visible in all chromospheric lines, but least prominent in Ca \textsc{ii} 8542 \AA\ line. In some pixels, broad asymmetric profiles with a single emission peak are produced, similar to the profiles observed in flare ribbons. They are caused by a deep onset of the chromospheric temperature rise {and large velocity gradients}.
}
{The simulation produces long fibrils similar to what is seen in observations. It also produces structures similar to flare ribbons despite the lack of non-thermal electrons in the simulation. The latter suggests that thermal conduction might be a significant agent in transporting flare energy to the chromosphere in addition to non-thermal electrons.}


\authorrunning{Bj{\o}rgen et~al.}

\maketitle

\section{Introduction}
\label{sec:intro}

{The chromosphere above active regions exhibits various dynamic phenomena. The most prominent features in quiescent periods of active region evolution are \textit{long fibrils}.} 
{Their existence was revealed already with the early solar observations that sampled \halpha\ line center \citep{1964PhDT........83B,1976SoPh...50...37M}.} 
{Later, with improved spectral and spatial resolution, a similar scene was found in other chromospheric lines such as Ca \textsc{ii} 8542 \AA\  and  \CaIIK\
\citep{2009A&A...502..647P,2011ApJ...742..119R,2019A&A...621A...1R}. }

{It is still, however not clear how they come about. Are they visible in chromospheric lines because of the density corrugations \citep{2012ApJ...749..136L} or do they outline magnetic tangential discontinuities where current sheets are formed \citep{2011ApJ...730L...4J}?}
{The most recently proposed scenario is that the long fibrils are actually contrails where hot material shoots up as a consequence of energetic events happening lower down \citep{2017A&A...597A.138R}.}
{It is also unclear if and how well  fibrils follow the magnetic field lines \citep{2011A&A...527L...8D,2017A&A...599A.133A,2013ApJ...768..111S}?} {Answering this question is crucial for chromospheric seismology \citep[e.g.][]{2012ApJ...750...51K} as well as techniques to constrain extrapolation of the magnetic field from
photospheric magnetograms \citep[e.g.][]{2016ApJ...826...61A}.}

{Another prominent feature in the chromosphere during the flaring phase in active regions, are \textit{flare ribbons}.} 
{According to the standard flare model \citep{2008LRSP....5....1B, 2011SSRv..159...19F}, these are locations where energy is deposited in the chromosphere predominantly by beams of energetic electrons precipitating down from the coronal reconnection site.}
{Strong chromospheric lines typically show broad emission peaks at those locations \citep{2015A&A...582A..50K, 2015ApJ...804...56R, 2018PASJ..tmp...61T,2018ApJ...861...62P} which models have difficulty to reproduce.}

{The main problem in modeling both long fibrils and flare ribbons is that the radiation originating from the solar chromosphere forms under complex non-Local Thermodynamic Equilibrium (non-LTE) conditions.}
{On the one hand, this makes  observations of the chromosphere difficult to interpret; on the other hand, the forward synthesis of chromospheric lines becomes quite complex and computationally expensive  \citep{2012ApJ...749..136L,2015ApJ...803...65S,2017ApJ...847..141S,2013ApJ...772...90L,2018A&A...611A..62B,2017A&A...597A..46S} if one wants to reproduce the spectral lines in detail. }

{The backbone of theoretical studies of 3D line formation in the solar chromosphere was so far  the "enhanced network" run \citep{2016A&A...585A...4C} computed with the 3D radiation-MHD code Bifrost \citep{2011A&A...531A.154G}.}
{A simulation including flux emergence performed with the same code has also been used \citep{2017ApJ...839...22H}.} 

{In case of flares, modeling of chromospheric lines has so far been limited to studies using 1D hydrodynamic models with heating by a prescribed electron beam  \citep[e.g.][]{2005ApJ...630..573A,2017ApJ...842...82R}.} 
{Line profiles from these simulations fail to reproduce the line widths.
However they can sometimes be achieved by adding artificial large velocity gradients to the simulation results.}
{Some observed profiles, typically in \MgIIHK, have a single emission peak with a very wide symmetric base. Such profiles cannot be simulated by velocity gradients or microturbulence only \citep{2017ApJ...842...82R}.}

{In this paper, we use a 3D MHD model that self-consistently shows both long fibrils and flare ribbons. It was computed using the radiation-MHD code MURaM \citep{2005A&A...429..335V,2017ApJ...834...10R}. 
The computational domain spans over} an entire active region, and contains a bipolar sunspot pair. The run produces a magnetic reconnection event in the simulated corona resembling observational characteristics of a flare and an eruption of cool chromospheric material \citep{2018NatAs.tmp..173C}.

The radiative transfer computations were performed in full 3D including partially coherent scattering (PRD) as well as charge conservation. 
{(which is particularly important for hydrogen and has a clear effect in the Ca \textsc{ii} and Mg \textsc{ii} intensities).} 
We investigate in detail the formation of the \CaIIK / $8542$ \AA, \MgIIk, and \halpha\ lines, all important diagnostics of the chromosphere.
We are motivated to do this because we want to test to what extent the MURaM simulation produces a chromosphere that resemble real observations, specifically, we are interested in whether the large spatial scales lead to long chromospheric fibrils, and how the coronal flare affects the chromosphere, despite the lack of non-thermal electrons in the model.

We compare the synthetic observations with observations taken with the Swedish 1-meter Solar Telescope
\citep[SST;]{2003SPIE.4853..341S} 
and the Interface Region Imaging Spectrograph
\citep{2014SoPh..289.2733D}. 

The paper is structured as follows: the modeling and methods are presented in Section \ref{sec:modeling},
an overview of the observations in Section \ref{sec:observation}, the results in Section \ref{sec:results}, and finally the
discussion and conclusions are given in Section \ref{sec:discussion}.

\section{Modeling}
\label{sec:modeling}

\subsection{Model atmosphere}
\label{sec:model_atmosphere}

\begin{figure*}
  \includegraphics[width = \textwidth]{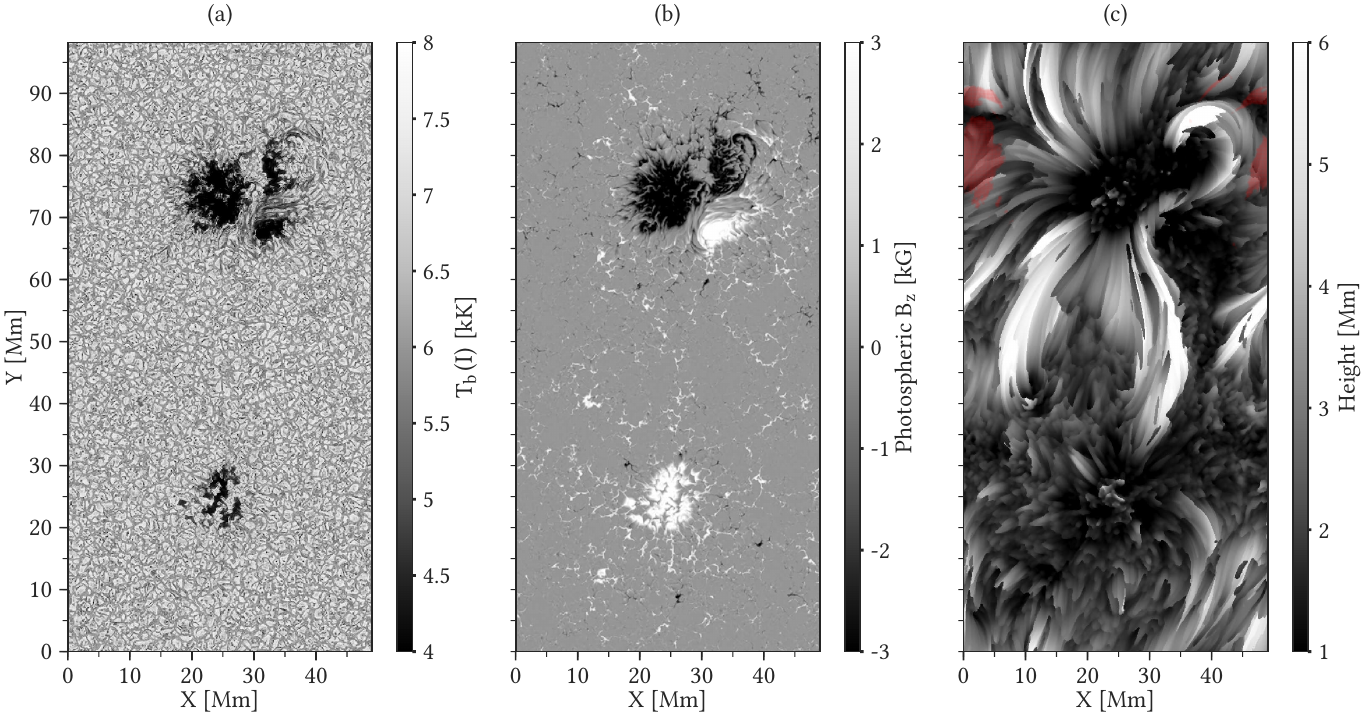}
  \caption{%
    Illustration of the model atmosphere.
    \emph{Panel a:} the vertically emergent intensity in the 5\,000\ \AA\ continuum.
    \emph{Panel b:} the vertical magnetic field strength where the optical depth at 5\,000 \AA\ is unity.
    \emph{Panel c:} the lowest height of the first occurrence of  a temperature of 0.5~MK.
    The zero point of the height scale is defined as the average height of where
    the optical depth at 5\,000 \AA\ is unity over a weakly-magnetized patch.
    The red areas in \emph{panel c} show where the CME arm is located in the original model atmosphere.%
    }
  \label{fig:atmos_photosphere}
\end{figure*}

{We ran a simulation with the same setup as in 
\citep{2018NatAs.tmp..173C},
but with double the spatial resolution.}
{We use a snapshot from this simulation taken at $71$ seconds before the flare peak, which is defined as the peak of the synthetic soft X-ray flux as it would be measured by the GOES satellite.}

The simulation spans in the vertical direction from $-$7.5~Mm below the photosphere to 41.6~Mm above it and includes the top of the convection zone, the photosphere, chromosphere and a part of the corona. The simulation box has $1024 \times 512 \times 1536$ grid points, which
correspond to $98.304~\text{Mm} \times 49.152~\text{Mm} \times 49.152~\text{Mm}$
in physical size, with a uniform horizontal grid of 96~km and a uniform
vertical grid of 32~km.
The simulation setup contains a bipolar active region that was present in the initial state. Near one of the sunspots a strongly twisted small bipolar magnetic structure was emerged by feeding in horizontal flux at the bottom boundary located about 7.5~Mm beneath the photosphere.

The simulation is run with gray LTE radiative transfer in the photosphere and chromosphere, optically thin radiative losses from the transition region and corona and included heat conduction along the magnetic field. The EOS uses a realistic solar composition and assumes local thermodynamic equilibrium. 

The primary goal of this numerical experiment was to simulate the evolution of coronal magnetic field self-consistently in 3D in a flare. For this reason, the processes relevant for chromospheric layers were not accounted for in detail. The gray approximation is chosen because it resulted in a chromospheric temperature stratification which agreed with 1D standard semi-empirical models of solar atmosphere much better than the standard LTE multi-group opacity scheme with four frequency bins
\citep{1982A&A...107....1N}.
Furthermore, the opacity and source function are set to zero at the base of the transition region, or the first grid cell with $T > 20$ kK, in order to avoid contributions from the corona in the downward directed rays. 

A slope-limited diffusion scheme was chosen so that overshoots are minimized in locations where the gradients can be significant, like near the flare ribbons. The problem is however not completely removed. For example, it results in cold pockets at the base of the transition region in some areas of the computational domain.
Finally, to relax the time-step constraints, the Boris correction and hyperbolic heat conduction are used.
For more details about the code, see
\citet{2017ApJ...834...10R}. 

Figure \ref{fig:atmos_photosphere} displays the model atmosphere. Panel (a)
shows the intensity in the continuum at 5\,000~\AA, with granulation and a pair of sunspot-like structures (but note the lack of penumbra around the spots), separated by 50~Mm. Panel (b) displays the vertical field strength, revealing that the upper sunspot is in fact a $\delta$-spot. In panel (c) we show the lowest height where the temperature is 50~kK, and can be thought of as a map of the height of the onset of the transition region. It shows elongated structures that reach up to 8~Mm and connect the sunspot pair.

In the model atmosphere, there is a long arm of material at chromospheric temperatures that has erupted up to a height of 18~Mm into the corona, reminiscent of a coronal mass ejection. For brevity we refer to this arm  as CME.

Since the CME has chromospheric temperatures, it should ideally be included in our radiative transfer calculations. We choose not to do so because this
would increase the computation time by a factor of 3, which we could not afford. Fortunately, the CME covers only 2.3\% of the surface area of the simulation domain. The surface area of the removed CME arm is shown as red patches in Fig. \ref{fig:atmos_photosphere}c.

We removed the upper part of the CME, so that the model atmosphere size is
reduced to $1024 \times 512 \times 290$ grid points, corresponding to heights
from $-$1.3~Mm beneath the photosphere up to 8~Mm above it.
The zero point of our height scale is defined as the height where the average optical depth at 5\,000 \AA\ is unity in a patch of the model atmosphere without strong vertical field strength. Due to the Wilson depression, there are columns where the height of optical depth unity at 5\,000 \AA\ is below the zero point of our height scale. The largest offset is $-$0.9~Mm located in the upper sunspot in Fig.~\ref{fig:atmos_photosphere}.

For our non-LTE calculations with hydrogen we resampled the vertical resolution of the model from 32~km to 16~km around the transition region to better resolve 
large gradients.
This resulted in a grid size of $1024 \times 512 \times 499$ points.
For magnesium and calcium, we kept the original grid spacing
because the spectra of these atoms are formed below the transition region.

In the MURaM simulation there exist locations where the temperature is as low as 1\,100~K.
Our background opacities are not correct at such low temperatures, and therefore we set the minimum temperature to 3\,250~K in the model atmosphere.
This has a minimal effect on the resulting spectra.

The simulation used a LTE equation of state.
The electron densities in the real chromosphere
are however far from LTE,
and should ideally be computed using non-equilibrium
ionization of at least hydrogen, and ideally also helium
\citep[e.g.,][]{2007A&A...473..625L}.
Computing the non-equilibrium ionization balance in post processing is not possible.
Therefore, we chose to compute the electron density from non-LTE statistical equilibrium for the contribution from hydrogen, and in LTE from all other elements {since they add a minimal number of electrons to the total in the upper chromosphere}. The details for this procedure are given in Section \ref{sec:linerized_eqs}.

\subsection{Radiative transfer computations}
\label{sec:radtrans}

We numerically solve the non-LTE radiative transfer problem with the Multi3D code
\citep{2009ASPC..415...87L} 
in a model atmosphere discretized on a Cartesian three-dimensional (3D) grid.
The code treats one atom in non-LTE by simultaneously
integrating the transfer equation and solving the statistical equilibrium
equations for the atomic level populations.
This is done iteratively using the multi-level accelerated $\Lambda$-iteration
(M-ALI) method with preconditioned radiative rates following
\citet{1991A&A...245..171R,  
          1992A&A...262..209R}. 
The transfer equation is integrated at spectral points covered by radiative
transitions of the atom using a domain-decomposed and parallelized method of
short characteristics
\citep[e.g.,][]{1987JQSRT..38..325O}.
For angle integrations we use a quadrature with 24 rays
\citep[set A4 from][]{carlson1963}. 
The opacities and the source function are approximated along the characteristics
using  linear interpolation, which is the most stable but the least accurate
method.
Our criterion for convergence is when the relative change in the populations is
${\leq}10^{-3}$.

In order to have a more accurate electron density, we first solve the non-LTE
problem for hydrogen, where the statistical equilibrium equations are
additionally constrained by the charge conservation equation (see
Section~\ref{sec:linerized_eqs}).
The obtained non-LTE electron density is re-used for solving the non-LTE problem
for \ion{Ca}{II} and \ion{Mg}{II}.

We treat the \ion{Ca}{II}~K and the \ion{Mg}{II}~k lines in partial
redistribution following the method by
\citet{2017A&A...597A..46S}. 
%

\subsection{Model atoms}
\label{sec:model-atom}

We used three different model atoms to compute the various
spectral lines from the model atmosphere.

To model \halpha\ we used a three-level plus continuum model atom of \ion{H}{I}, that was constructed by removing the $n=4$ and $n=5$ levels from the atom used by
\citet{2012ApJ...749..136L}.
Following them, we treated \lyalpha\ in complete redistribution (CRD) with a Doppler absorption profile, to avoid the large computational cost of treating this line in partial redistribution (PRD).

First, we experimented with the original five-level plus continuum model atom,
which produced population inversions in the Ly$\gamma$ and Ly$\delta$ transitions
at several places of the model atmosphere.
Such inversions cause negative opacities as well as negative source functions.
Although the analytical formal solution of the transfer equation allows this,
such inversions often produce numerical instabilities, especially, in the method
of short characteristics.
Algorithms used for interpolating the source function and the opacity either
to facets of grid voxels or along the short characteristic, have inherent
numerical errors, which might produce at the same time opposite signs of the
source function and optical depth. Such an unphysical situation destroys the convergence of the M-ALI iteration
scheme.

We tried the method of
\citet{2002A&A...384..562K}, 
who proposed to branch the integration of the transfer equation in terms of
emissivities and opacities instead of source functions and optical depths, but
found this to be unstable too.
Finally, we simply removed the upper levels of the transitions that had population
inversions from the model atom.

For \MgII\ we used a four-level plus continuum model atom of
\citet{2013ApJ...772...89L},
where we treat only the \MgII~k line in PRD and \MgII~h in CRD to lower the computational costs.

Finally, for \CaII\ we used the five-level plus continuum
model atom of
\citet{2018A&A...611A..62B}
where we similarly treat only \CaII~K in PRD and \CaII~H in CRD.

We tested using the 1D FAL-C model atmosphere
\citep{1993ApJ...406..319F} 
to see whether the neglection of PRD in one of the doublet's transitions 
changes the emergent intensity in the other line.
In \ion{Ca}{II}~K only the inner core was marginally affected.
In \ion{Mg}{II}~k the core was changed by 1.7\% and the emission peaks were
changed by less than 8.5\%.
In each doublet, \ion{Ca}{II} K and H as well as \ion{Mg}{II} k and h, both
lines share similar formation properties, so in the following we focus on the
\ion{Ca}{II}~K and the \ion{Mg}{II}~k lines only.

The emergent intensities as well as the formation heights slightly change owing
to all these simplifications in our model atoms.
These changes are relatively small. Since we are not concerned about 
detailed properties, they do not change the overall results and conclusions of the paper.
\subsection{Charge conservation}
\label{sec:linerized_eqs}

A correct electron density is important for non-LTE modeling of chromospheric lines because the collisions with electrons affect the coupling of the source function to the local conditions of the gas.
The model atmosphere does not explicitly contain the electron density, but  MURaM implicitly assumes LTE electron densities through its equation-of-state.
LTE is however a bad approximation of the electron density in the chromosphere and transition region, and the electron density should ideally be computed including non-equilibrium ionization of both hydrogen and helium
\citep{2002ApJ...572..626C,2007A&A...473..625L,2016ApJ...817..125G}.

Computing non-equilibrium ionization in post-processing is not possible, so
instead we derive the non-LTE electron density
through solving the non-LTE problem for hydrogen together with the charge
conservation equation.

First, we tried an approach similar to what the RH code uses
\citep{2001ApJ...557..389U}.
The electron density is updated after every M-ALI iteration to be consistent
with the proton density, while the ionization of all other elements is computed
in LTE.
We found this approach to unstable in our calculations and the NLTE populations for hydrogen did not converge for our problem.

Instead, we decided to solve simultaneously in each M-ALI iteration the equation
of charge conservation and the system of statistical equilibrium equations for
preconditioned rates.
This new system of equations is non-linear with respect to the unknown electron
density and hydrogen level populations as some of the rates depend on the
electron density.
We solve it iteratively using the Newton-Raphson method.
A similar technique has been used in other radiative transfer codes
\citep[e.g.,][]{1992ApJ...397L..59C,1995A&A...302..587P,1995A&A...299..563H}. 
Our implementation is explained in Appendix~\ref{app:charge_conservation}.

We note that this is a substantial improvement over previous radiative transfer 
calculations with the Multi3D code for hydrogen, where the proton density could be larger than the electron density 
\citep{2012ApJ...749..136L,2017ApJ...839...22H}.
%

\subsection{Broadening of H$\alpha$} \label{sec:broadening_halpha}

The default Stark broadening mechanism for H$\alpha$ in Multi3D is based on the
approximation of
\citet{1978JQSRT..20..333S}. 
When the electron density exceeds $10^{13}$~cm$^{-3}$ in
the chromosphere, this approximation becomes inaccurate.
We tested whether we should include  the more accurate unified Stark broadening theory as recently proposed by
\citet{2017ApJ...837..125K}
to model the line profile.
We performed 1D tests with several columns from the model atmosphere.
For columns with chromospheric electron densities above $10^{13}$~cm$^{-3}$ we
indeed found that the line profile computed with the unified theory broadens more.
However, most profiles do not show a substantial difference since the typical
electron densities are not as high in the line formation region.
Since the unified theory is computationally expensive owing to a convolution
operation, we decided not to include the unified Stark
broadening theory for H$\alpha$.

\subsection{\halpha\ in 1D and 3D} \label{sec:broadening_halpha}

\begin{figure}
  \includegraphics[%
    width = \columnwidth,%
    trim = 0pt 0pt 0pt 0pt,%
    clip = true%
  ]{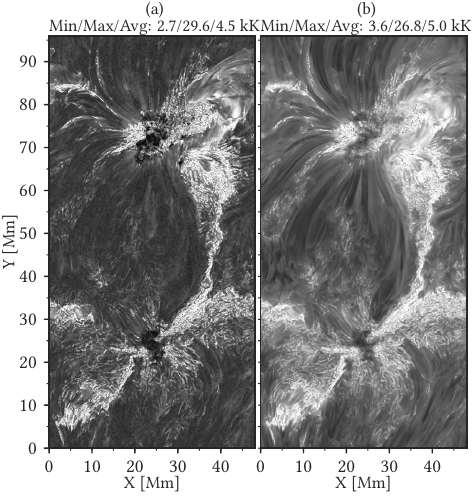}%
  \caption{%
    Simulated images of \halpha\ at nominal line center, $\Delta \lambda = 0$ \AA , from the model atmosphere in vertically emergent intensity. 
    The images are given in brightness temperature $T_\text{b}$ and are clipped
    at  2.8 and 8.5~kK.
    \textit{Panel (a)}: 1.5D radiative transfer where each column in the snapshot is treated as a plane-parallel atmosphere.
    \textit{Panel (b)}: full 3D radiative transfer.
    The minimum, maximum, and average $T_b$ for the entire field-of-view is provided in a label above each panel.
    }
  \label{fig:halpha_1d_vs_3d}
\end{figure}

Previous \halpha\ studies using a 3D evaluation of the radiation field were performed using quiet-Sun conditions 
\citep{2012ApJ...749..136L}.
Under those circumstances, the intensities computed in 1.5D, where the effect of horizontal scattering is neglected, showed a strong imprint of photospheric granulation. A 3D evaluation of the radiation field was needed to obtain chromospheric features in the emerging intensities. In the present study, we are analyzing an active region, and we wanted to assess the importance of horizontal scattering compared to the quiet-Sun case.

Figure \ref{fig:halpha_1d_vs_3d} shows the comparison between the vertically emergent intensity in 1D and 3D using our active region snapshot. The 1D computation shows chromospheric structures and traces of the large-scale structures, but much less enhanced than in the 3D case. We note that an imprint of the photosphere is still visible through the fibril-like structures in the 1D case, but not as prominent as in the quiet-Sun case. 

The brightest features have a similar morphology in the 1D and 3D computations. Analysis shows that this is caused by the high mass density, and thus strong collisional coupling of the source function to the temperature, in the chromosphere at these locations (see  Sec.~\ref{sec:flare_profiles}). Therefore, this can potentially open the possibility to include \halpha\ in inversions that assume 1D geometry
\citep{2018A&A...617A..24M,2018arXiv181008441D} 
in active regions.

\section{Observations}
\label{sec:observation}

\begin{figure*}
  \includegraphics[width = \textwidth]{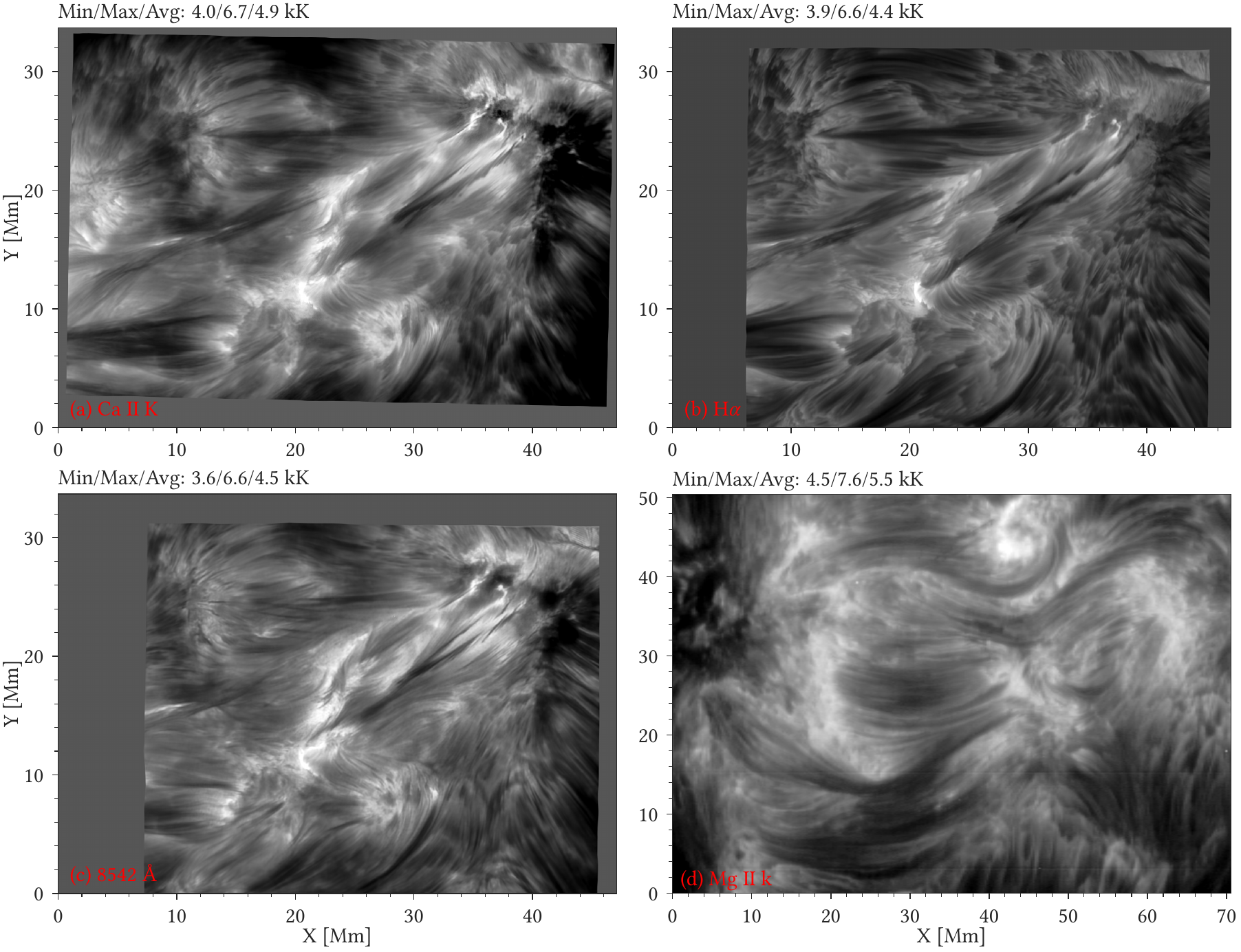}
  \caption{%
    Observations of two active regions close to the disk center, at the nominal
    line center of
      \emph{a)} \ion{Ca}{II}~K (SST/CHROMIS),
      \emph{b)} H$\alpha$ (SST/CRISP),
      \emph{c)} \ion{Ca}{II} 8\,542~\AA\ (SST/CRISP), and
      \emph{d)} \ion{Mg}{II}~k (IRIS).
    \emph{Panels a--c:} NOAA 12593.
    \emph{Panel d:} NOAA 12494.
    The images are given in brightness temperature $T_\text{b}$ and are clipped
    at 4.5--6~kK for \CaIIK, 3.9--6~Kk for \halpha, 3.8--6~kK for 8\,542 \AA,
    and 4.7--7~kK for \MgIIk.
    Above the top left corner of each panel we show the minimum, maximum, and
    average brightness temperature for the FOV.
    \emph{Panels a--c} have the same coordinate scale, while \emph{panel d} has a different scale.%
   }
  \label{fig:observations}
\end{figure*}

We use two different observational datasets to compare with the synthetic data.
The first dataset is an observation of active region NOAA 12593, 
located near to the disk center at $\theta_x= -68$\arcsec and
$\theta_y = -6$\arcsec (helioprojective-Cartesian coordinates), taken at the
Swedish 1-meter Solar Telescope
\citep[SST;]{2003SPIE.4853..341S} 
{ on September 19, 2016, at 09:31--09:57~UT.}

The data was recorded simultaneously by the CHROMIS and CRISP
\citep{2008ApJ...689L..69S}
instruments, which observed Fe \textsc{i} 6301 \AA,
Fe \textsc{i} 6032 \AA,  \halpha, Ca \textsc{ii} 8542 \AA, and
Ca \textsc{ii} K. We only use the latter three spectral lines.
The data were processed with the CRISPRED and CHROMISRED pipelines
\citep{2015A&A...573A..40D,
          2018arXiv180403030L}.
For a more detailed description of the SST observation we refer to
\citet{2018A&A...612A..28L},
here it suffices to mention that we obtained near-simultaneous spectral scans in
all three lines.

Unfortunately, there is no co-observation with the Interface Region Imaging Spectrograph
\citep[IRIS][]{2014SoPh..289.2733D}. 
Therefore, we use IRIS data observed on February 5, 2016, at 08:21--09:12~UT in order to be able to compare our computations of the \MgIIk\ line.
The IRIS target is an emerging active region NOAA 12494, observed near the disk center at $\theta_x =  -96$\arcsec and $\theta_y = -98$\arcsec.

Figure \ref{fig:observations} shows the line core images for all four spectral lines.
Panels~a-c show an active region covered with fibrils and a sunspot in the upper right
corner.
Panel~d shows a slit jaw-image at the nominal line center of \ion{Mg}{II}~k,
of an active region with one sunspot located on the left side and a sunspot outside
the FOV on the right side.
Long elongated fibrils are connected with the sunspots and pores
located in the middle of the FOV.

\section{Results} \label{sec:results}
\subsection{The effect of non-LTE electron density on the emergent intensity}
\label{sec:ccem}
%
\begin{figure*}
  \includegraphics[width = \textwidth]{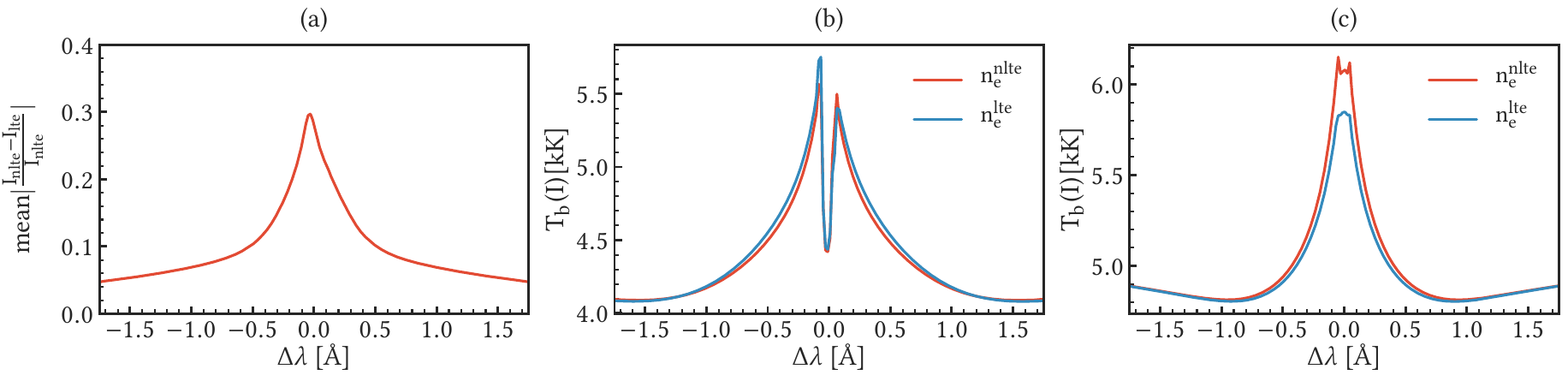}
  \caption{%
    Dependence of the vertically emergent intensity in \ion{Mg}{II}~k on how the
    electron density is computed.
    \emph{Panel a}: spatially-averaged relative change in the intensity, which
    is computed using either non-LTE or LTE electron densities.
    \emph{Panels b,c}: line intensities expressed as brightness temperature
    $T_\text{b}$ and computed using LTE (blue) or non-LTE (red) electron density
    at two structures indicated in Fig.\ \ref{fig:fig3_height}.
      \emph{Panel b}: a dark fibril, red dot in Fig.\ \ref{fig:fig3_height}.
      \emph{Panel c}: a flare ribbon, blue dot in Fig.\ \ref{fig:fig3_height}.
  }
  \label{fig:nlte_lte_edens}
\end{figure*}

We compared how different are electron densities computed either using
the LTE equation-of-state for all chemical species or by treating hydrogen in
non-LTE with imposed conservation of charge.
We found severe differences between the LTE and non-LTE values in the middle and upper chromosphere.

On average the {inclusion of} charge conservation {in the calculations,} leads to higher electron densities {compared to LTE} in cold areas of the chromosphere, owing to an increased hydrogen ionization degree through the Balmer continuum. This difference was largest (by a factor $\sim 10^3$) in cold pockets {(3250 K)} right below the transition region. {In hot areas in the mid chromosphere we typically find a somewhat lower electron density than in LTE.}

Figure~\ref{fig:nlte_lte_edens} illustrates how the non-LTE electron densities
change the emergent intensity in \MgII~k.
For the spatially-averaged intensity, the maximum relative difference
reaches 30\% near the line core (Fig.\ \ref{fig:nlte_lte_edens}a).
Individual profiles show even larger differences depending on which
chromospheric structure they are emerging from.
In dark fibrils the non-LTE intensity drops below the LTE one
(Fig.\ \ref{fig:nlte_lte_edens}b), but in flare ribbons   it increases
significantly near the line core (Fig.\ \ref{fig:nlte_lte_edens}c).

Therefore, in the rest of the paper we only present results using the
non-LTE electron density.

\subsection{Formation heights}
\label{sec:formation_height}
%
\begin{figure}
  \includegraphics[%
    width = \columnwidth,%
    trim = 0pt 0pt 0pt 0pt,%
    clip = true%
  ]{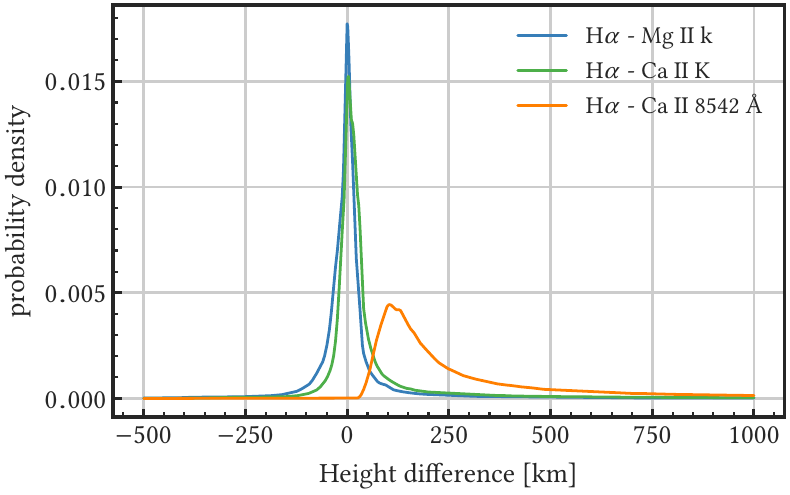}%
  \caption{%
    Distribution of the differences between the line-core formation heights of
    H$\alpha$ and, respectively, \MgIIk\ (blue), \CaIIK\ (green), and
    \ion{Ca}{II} 8\,542~\AA\ (orange).
  }
  \label{fig:formation_height}
\end{figure}
\begin{figure}
  \includegraphics[%
    width = \columnwidth,%
    trim = 2pt 2pt 2pt 2pt,%
    clip = true%
  ]{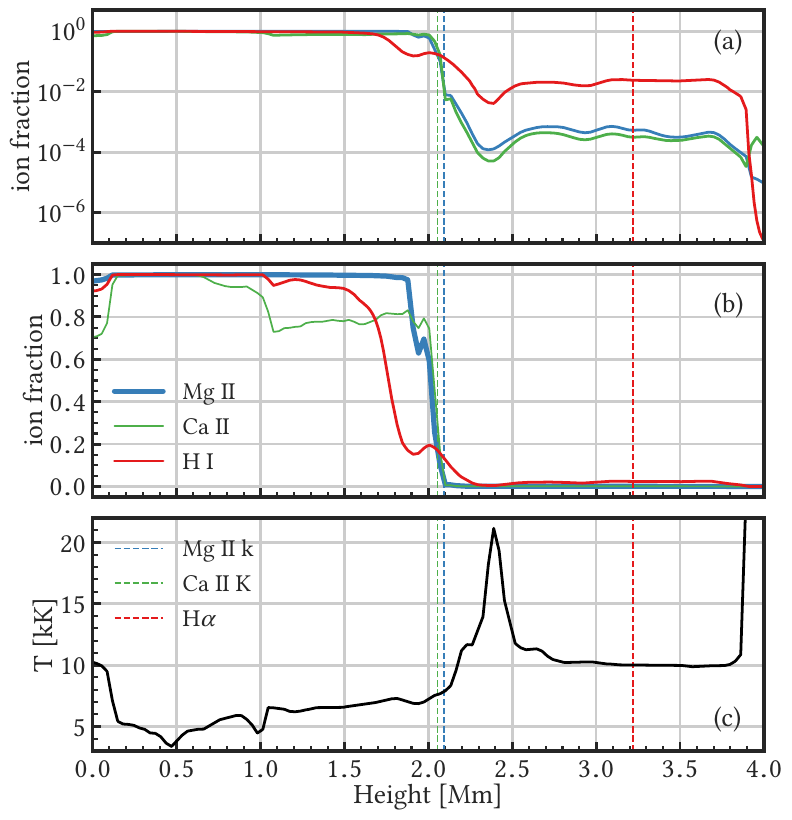}%
  \caption{%
     {Fraction of an element in a given ionisation state} for \ion{Mg}{II} (blue solid), \ion{Ca}{II} (green
    solid), and \ion{H}{I} (red solid) in a  column of the MURaM model
    atmosphere.
    \emph{Panel a}: log-scaled ionization fractions.
    \emph{Panel b}: linearly-scaled ionization fractions.
    \emph{Panel c}: gas temperature (black solid) as a function of height.
    Vertical lines show the maximum formation heights of \ion{Mg}{II}~k (dashed
    blue), \ion{Ca}{II}~K (dashed green), and H$\alpha$ (dashed red).
  }
  \label{fig:formation_height2}
\end{figure}

Calcium and magnesium are both predominantly singly ionized in the chromosphere.
\ion{Mg}{II} ionizes to \ion{Mg}{III} at slightly higher
temperatures owing to the higher ionization potential of \MgII\ (15.0 eV)
compared to \CaII\ (11.87~eV).
Both the  \MgIIHK\ and \CaIIHK\ lines have the ground state as their lower
level, so their opacity scales approximately with
the density as long as \MgII\ and \CaII\ are the dominant ionization stage.
As \ion{Mg}{II} is 18 times more abundant than \ion{Ca}{II}, the
line cores of \MgIIHK\ are formed a few scale heights higher up in the
chromosphere than the line cores of \CaIIHK\
\citep{2013ApJ...772...90L,2018A&A...611A..62B}. 

In contrast, the lower level of \halpha\ is an excited state at
10.2~eV above the ground state.
In LTE this makes the line opacity very sensitive to the temperature.
In the solar chromosphere, the H$\alpha$ line opacity depends in a
complicated way on the temperature, density, and the time history of
the atmosphere owing to the non-equilibrium ionization of hydrogen.
\citet{2012ApJ...749..136L}
investigated the formation of \halpha\ in a radiation-MHD simulation of
the quiet Sun
and found that in those circumstances the line can be reasonably accurately
modeled assuming statistical equilibrium.
In that simulation they also found that
the line opacity weakly depended on temperature and mainly
depended on mass density.
\citet{2012A&A...539A..39C} 
investigated the ionization of \HI, \CaII, and \MgII, and
found that at temperatures above 15~kK, the fraction
of \HI\ is much larger than the fractions of
\CaII\ and \MgII. 
\citet{2017A&A...597A.138R}
and
\citet{2017A&A...598A..89R}
argued similarly, albeit based on LTE, that \halpha\ can have an appreciable
opacity at temperatures well above 20~kK.

Studies of the formation heights of the
line cores based on the quiet-Sun-like
simulation of
\citet{2016A&A...585A...4C}
indicate that \halpha\ forms 300--1000~km below
\CaIIK\ and \MgIIk\ 
\citep{2013ApJ...772...90L}.
The similarity of the line-core images in
Fig.~\ref{fig:observations}a,b
casts doubt on the validity of this result in active regions.
We therefore investigated the formation heights in the MURaM simulation.

We measured the formation heights at optical depth unity in the
corresponding line cores of \halpha, \MgIIk, \CaIIK, and \ion{Ca}{II} 8\,542~\AA.
Figure~\ref{fig:formation_height} shows
differences between the formation height of \halpha\ and the formation
heights of the rest of the lines.
On average, the core of H$\alpha$, k$_3$, and K$_3$ features are formed
at similar heights and {roughly 150~km} above the core of 8\,542 \AA\ throughout the
field-of-view.

We note that the distributions have long tails towards
positive values meaning that \halpha\ {can be formed}
much higher than the other lines.
These points, where the height difference exceeds 1\,000~km,
constitute for the field-of-view
${\sim}$2\% for both \MgIIk\ and \CaIIK, and $\sim$5\% for 8\,542\ \AA.
Figure \ref{fig:formation_height2} illustrates this case.
At 3.22~Mm where the \halpha\ core reaches optical depth unity, \ion{H}{I}
is only 97.5\% ionized and at the same location the metals
(\CaII\ and \MgII) are 99.9\% ionized.
\halpha\ is formed highest in the atmosphere.  {It retains opacity where the lines from the metals have none because \HI\ does not ionize as quickly as \CaII\ and \MgII\ when the temperature increases.}

\subsection{Synthetic images }
\label{sec:synthetic_images}
%
\begin{figure*}
  \includegraphics[width = \textwidth]{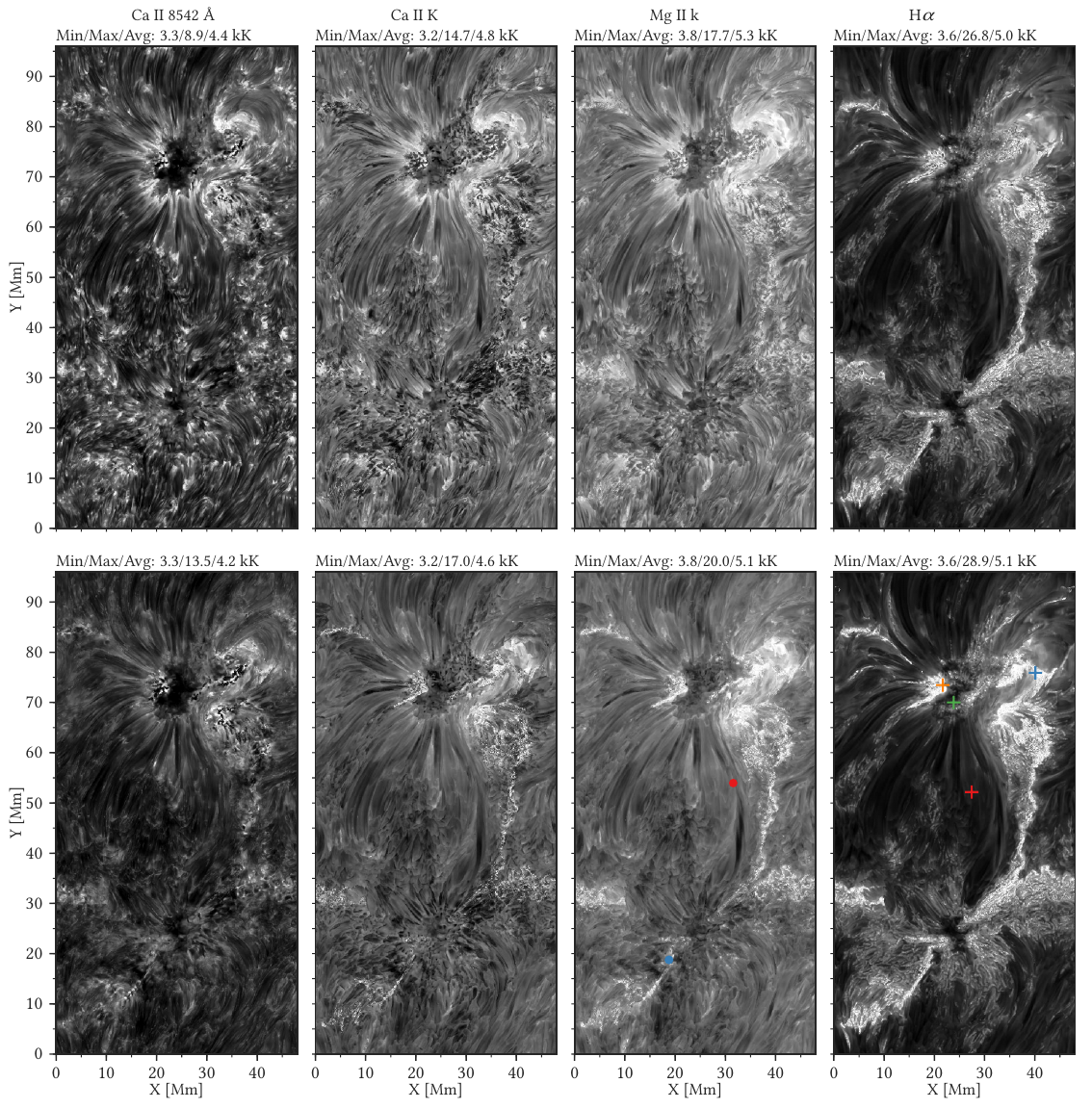}
  \caption{%
    Simulated images of \ion{Ca}{II}~8\,542~\AA, \CaIIK, \MgIIk, and \halpha\
    from the model atmosphere in vertically emergent intensity.
    \emph{Upper row}: at nominal line center, $\Delta\lambda = 0$~\AA.
    \emph{Bottow row}: at wavelengths corresponding to the maximum formation
    height, $\max z(\tau = 1)$.
    Intensity is expressed as brightness temperature $T_\text{b}$ and is clipped
    to same ranges in each column: 3.6--7~kK for \ion{Ca}{II}~8\,542~\AA, 3.6--7~kK
    for \CaIIK, 3.6--8~kK for \MgIIk, and 3.6--10~kK for \halpha.
    The minimum, maximum, and average $T_\text{b}$ for the entire field-of-view
    is provided in a label above each panel.
    Colored symbols mark the following structures:
      a dark fibril (red dot) and a flare ribbon (blue dot),
        see Fig.\ \ref{fig:nlte_lte_edens};
      an active region in emission (orange cross),
        see Fig.\ \ref{fig:flare_profiles};
      an active region in emission with a strong velocity gradient (blue cross),
        see Fig.\ \ref{fig:flare_profiles_2};
      a sunspot (green cross), see Fig.\ \ref{fig:sunspot_profiles};
      a fibril (red cross), see Fig.\ \ref{fig:fibril_profiles}.
  }
  \label{fig:fig3_height}
\end{figure*}

In Fig.~\ref{fig:fig3_height} we show the vertically emergent intensity in the line core of \CaII\ 8542~\AA, \CaIIK, \MgIIk, and \halpha.
We first describe what we see at nominal line center
(Fig.\ \ref{fig:fig3_height}, top panels).
In all four spectral lines we see large elongated structures
emanating from the two sunspots (see
Fig.\ \ref{fig:atmos_photosphere}b),
similar to large-scale fibrils seen in observations.
From now on we call the structures on our synthetic images fibrils too.
The fibrils are much more pronounced in \halpha\ than in the other lines.
The longest fibrils extend up to 35~Mm.

Careful comparison of the panels reveals that the fibrils appear
somewhat different in \ion{Ca}{II} 8\,542~\AA, looking translucent with the
background chromosphere shining through.
This is especially visible in the upper-right corner.
This is consistent with the lower formation height of this line, as discussed in Section~\ref{sec:formation_height}.

In H$\alpha$,
we observe several very bright patches close to the sunspots, which connect
with a thinner bright arc.
They look similar to flare ribbons.
Along these ribbons the \halpha\ profiles are in emission and very broad.
In \CaIIK\ and \MgIIk\ we see
only a weak signature of the ribbon, while in \ion{Ca}{II}
8\,542~\AA\ it is only visible as a small brightness enhancement
to the right of the upper sunspot. 

Now we describe what we see at wavelengths where the intensity in each pixel individually is taken at the wavelength where the formation height is largest. (Fig.\ \ref{fig:fig3_height}, bottom panels).
These images thus compensate for Doppler shifts of the line core
\citep{2013ApJ...772...90L,2018A&A...611A..62B}.

The H$\alpha$ image is relatively unchanged, while the other images are more
affected.
This difference is caused by the large thermal broadening of
H$\alpha$ compared to the other lines, which makes it less sensitive to
Doppler motions.
In the \ion{Ca}{II} and \ion{Mg}{II} lines the carpet of fibrils now
appears thicker and sharper than in the nominal line-core images.
The flare ribbons are now visible in these lines, 
even in \CaII~8\,542~\AA .

The simulated images exhibit features that quite resemble observational
features in our Fig.\ \ref{fig:observations} or
Fig.\ 1 by
\citet[][]{2011ApJ...742..119R}.
The observations show that \CaIIK\ and H$\alpha$ trace roughly the same structures.
The \CaII~8\,542~\AA\ often shows different fibrils, with more visible
patterns of waves and shocks from the lower chromosphere.

The SST/CRISP, SST/CROMIS, and, to a lesser extent, IRIS observations 
show finer structures than the synthetic images.
This is most likely caused by the finite horizontal grid spacing (96~km) of the
model atmosphere compared to the resolution of the observations ($\sim$100~km
for SST, $\sim$250~km for IRIS).
\citet{2018A&A...611A..62B} 
showed for radiation-MHD models of the enhanced network that synthetic
fibrils get thinner and closer in appearance to the observed ones
if the horizontal grid spacing of the simulation decreases from $48$ km to $31$ km.

\citet{2012ApJ...749..136L,
          2013ApJ...772...90L}
and
\citet{2018A&A...611A..62B} 
computed synthetic images of the chromosphere in \halpha, \CaIIK,
and \MgIIk\ from the quiet Sun simulation by
\citet{2016A&A...585A...4C}.
Their line-core images showed ordered fibrils stretching between two opposite
magnetic polarities separated by about 8~Mm, which also limited the fibril
length to 8~Mm.
\citet{2017ApJ...839...22H}
computed H$\alpha$ and \MgIIh\ for a 
flux emergence simulation, also computed with the Bifrost code.
Their H$\alpha$ line-core image showed some thick unordered
fibrils with a maximum length of 12~Mm, while
the corresponding \MgIIh\ image does not show
 a clear fibril structure.

The current MURaM simulation is the first simulation including a chromosphere that contains the
spatial scales of an entire active region, and succeeds in reproducing long fibrils connecting the opposite polarities
of an active region as seen in observations.

In the following subsection we analyse some of the structures seen in the synthetic intensity images in more detail.

\subsection{Synthetic fibrils}
\label{sec:synthetic_fibrils}
%
\begin{figure*}
  \includegraphics[width = \textwidth]{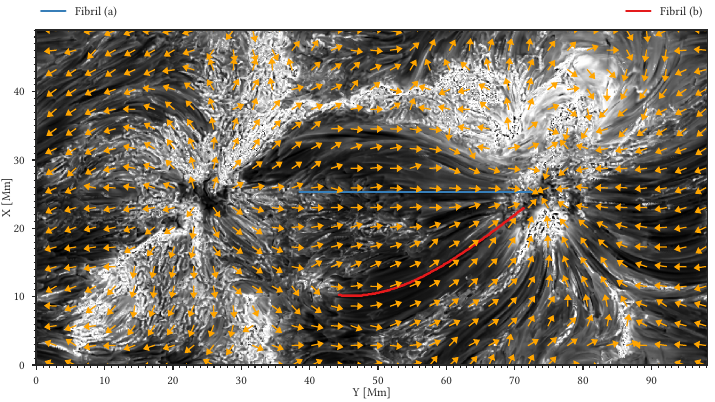}
  \caption{%
    Image of \halpha\ at nominal line center overplotted with the 
    horizontal magnetic field directions (orange arrows) at the formation height. 
    The arrows are plotted every 30th grid point.   
    The visibility of H$\alpha$ fibrils is artificially enhanced using unsharp
    masking.%
    The blue and red lines indicate two slices along selected fibrils,
    illustrated in Fig.\ \ref{fig:fibrils} (blue) and
    Fig.\ \ref{fig:fibrils_magnetic} (blue and red).
  }
  \label{fig:overview_azimuth}
\end{figure*}
\begin{figure*}
  \includegraphics[width = \textwidth]{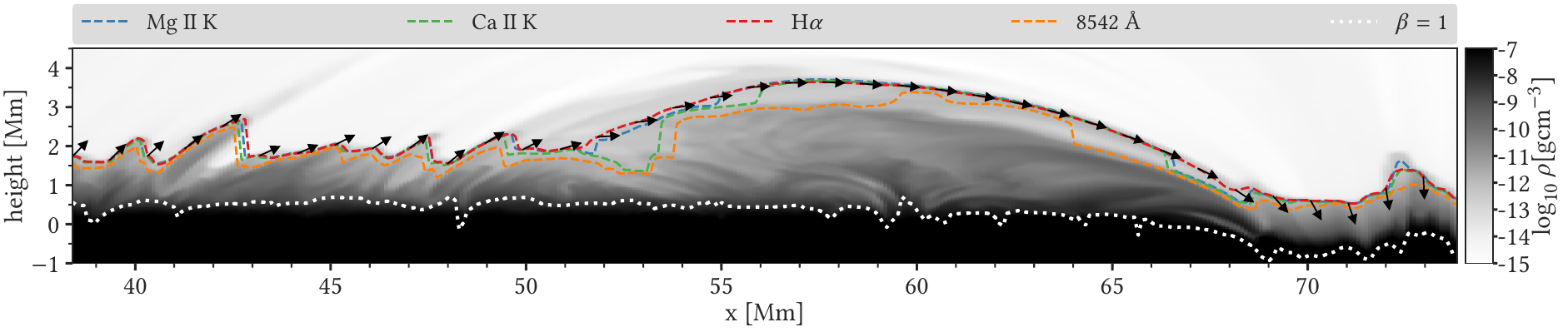}%
  \caption{%
    Mass density in a vertical cut through the model atmosphere along the blue
    line in Fig.~\ref{fig:overview_azimuth}.
    For each spectral line mentioned in the legend, corresponding dashed curves
    show the maximum formation height at optical depth unity.
    The white dotted line shows the first occurrence from the upper convection zone where the magnetic pressure equals gas pressure ($\beta=1$).
    The arrows show the direction of the magnetic field vector in the plane of the cut along the formation height curve of \halpha. They are plotted every tenth grid point.
  }
  \label{fig:fibrils}
\end{figure*}
\begin{figure*}
  \includegraphics[width = \textwidth]{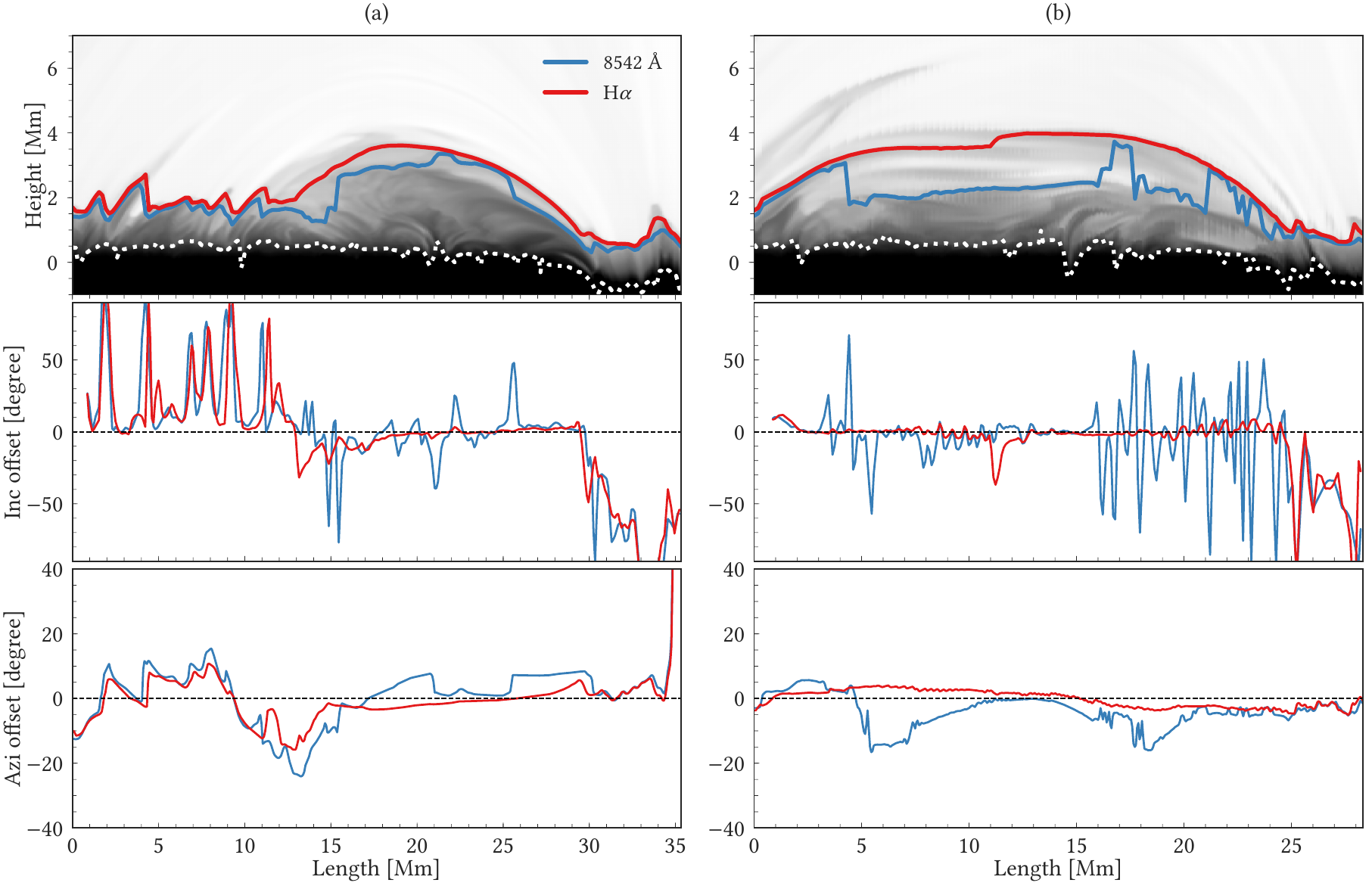}%
  \caption{%
    Alignment between the magnetic field vector and the curve of the maximum
    formation height at optical depth unity in two selected vertical slices.
    \emph{Left column a}: the blue curve fibril in Fig.\ \ref{fig:overview_azimuth}.
    \emph{Right column b}: the red curve fibril in Fig.\ \ref{fig:overview_azimuth}.
    \emph{Top row}: formation heights of H$\alpha$ (red) and \ion{Ca}{II}
    8\,542~\AA\ (blue) with the $\beta = 1$ height (white dotted) layered over
    the mass density (grayscale background).
    \emph{Middle row}: the angle between the inclination of the magnetic field and the $\tau=1$ curve.
    \emph{Bottom row}: the angle between the magnetic field azimuth and the azimuth of the fibril.
  }
  \label{fig:fibrils_magnetic}
\end{figure*}

The horizontal domain size of the current MURaM simulation is 98~Mm${}\times{}$49~Mm.
This large size allows for $\sim$40~Mm long magnetic loops in the chromosphere
that connect the two sunspots.
The synthetic images in Fig.\ \ref{fig:fig3_height} show long fibrils that
follow the large-scale structure of the field.
We investigated whether the fibrils trace the magnetic field, both the vertical
and horizontal components.

Figure \ref{fig:overview_azimuth} shows the line-core intensity of \halpha, with
the horizontal component of the magnetic field at the formation height overplotted.
The fibrils follow the horizontal magnetic field direction in most places, but
at some locations (such as the arrow at $(X, Y) = (54, 34)$~Mm)
they do not.

Figure~\ref{fig:fibrils} illustrates the magnetic field component in the plane
of a model atmosphere slice that follows the blue line in Fig.~\ref{fig:overview_azimuth}.
{As expected for active regions,} the magnetic pressure exceeds the gas pressure
at low heights in the atmosphere, as illustrated by the plasma
$\beta = 1 = P_\text{gas} / P_\text{B}$ curve, which goes from $-$1~Mm to
$0.5$~Mm.
The lowest height of $\beta = 1$ curve is located at the
sunspot ($X > 69$~Mm), where the magnetic field strength reaches 6~kG in the
photosphere.
Horizontally the fibril spans from $X = 51$~Mm to $X = 68$~Mm and for $X < 51$~Mm the cut intersects an area
of the atmosphere that shows irregular short structures in the H$\alpha$
image of Fig.~\ref{fig:overview_azimuth}.

In the sunspot, the transition region is very low, 1~Mm above the quiet photosphere {and roughly 2~Mm above the local height of continuum optical depth unity}. so that all four spectral lines are formed mostly at the same height.
The angle between the field vector and the $\tau=1$ surface is large.
In the area  where $x < 51$~Mm we see intrusions of high
mass density sticking into the low-density corona, and the $\tau=1$ surface follows these mass intrusions.
We speculate that these could be Type~I spicules as simulated before in 2D 
\citep[e.g.,][]{2006ApJ...647L..73H},
but this is impossible to confirm without analyzing a time series.

Now we turn our attention to the long loop-like structure between $X = 51$~Mm and
$X = 68$~Mm {in Figure~\ref{fig:fibrils}.}
The magnetic field vector and the $\tau=1$ height curve of H$\alpha$ are parallel to each other for
a large fraction of the loop.
It is not shown in the figure, but this is true to a lesser extent for the other lines, which all exhibit jumps in their formation height.
We analyze a number of other fibrils not illustrated here and found that for clearly defined fibrils, H$\alpha$ indeed tends to follow the magnetic field lines.
This is in contrast with the results from 
\citet{2015ApJ...802..136L},
who found in a simulation of quiet sun fibrils that H$\alpha$ fibrils
typically follow the horizontal component, but not the vertical component, of the magnetic field.

To investigate the magnetic field we traced two fibrils to get a more detailed overview. They are marked by
blue and red curves in Fig.\ \ref{fig:overview_azimuth}.
Figure \ref{fig:fibrils_magnetic} shows the offset between the inclination of the $\tau=1$ curve and magnetic field vector, and the offset between the azimuths of the fibril and magnetic field.
We only show results for H$\alpha$ and \CaII~8\,542~\AA, as they are the lines with the highest and lowest formation heights.

Figure~\ref{fig:fibrils_magnetic}a shows the same fibril as in Fig.~\ref{fig:fibrils}
inbetween 51~Mm and 68~Mm.
The inclination offset is around $10^\circ$ for both lines.
However, the formation height curve of H$\alpha$ is more smooth, while the \CaII~8\,542~\AA\ curve has a number of
sudden jumps, which cause spikes in the offset curve.
Physically, these jumps mean that depending on where you look along
the fibril, you see different field lines, but locally the formation height curve
follows the magnetic field inclination.
The H$\alpha$ fibril follows the azimuth very well, while the
\CaII~8\,542~\AA\ shows offsets up to 10\degr.

Figure \ref{fig:fibrils_magnetic}b shows the offsets of a curved fibril indicated with a 
red line in Fig.~\ref{fig:overview_azimuth}.
We see almost the same behavior as the previous fibril shows: in H$\alpha$ it
follows both the inclination and the azimuth of the magnetic field vector very well,
while in \CaII~8542~\AA\ it does it less so owing to large changes in the formation height. 

In addition to the fibrils analyzed here we manually investigated a number of other fibrils, and they show similar behavior.
We thus conclude that most fibrils in this active region simulation do trace the horizontal magnetic field, especially for spectral lines formed in the upper chromosphere (\CaIIHK, \MgIIHK, and \halpha).
Fibrils seen in 8\,542 \AA\ mostly, but not always, trace the horizontal component of the magnetic field.
This agrees with the observational studies
\citep{2011A&A...527L...8D,2017A&A...599A.133A}.
We note that this simulation does not include ambipolar diffusion.
Inclusion of this process allows the magnetic field to be less
aligned with density and temperature structures in the chromosphere because it
relaxes the frozen-in condition
\citep{2016ApJ...831L...1M}.
However, simulations with ambipolar diffusion have only been performed in 2.5D simulations \citep{2012ApJ...750....6C} or simplified 3D simulations 
\citep{2007ApJ...666..541A} or radiative 3D simulation with only photosphere  \citep{2017A&A...601A.122D}.
Therefore, only a full 3D MHD simulation that includes chromosphere with ambipolar diffusion should
be used to investigate the possible misalignment properly 
\citep[see also Sec.~3.8.3 of ][]{2014LRSP...11....3C}. 
%
\subsection{Spatially averaged spectral profiles}
\label{sec:spatially_averaged}
%
\begin{figure}
  \includegraphics[%
    width = \columnwidth,%
    trim = 0pt 0pt 0pt 0pt,%
    clip = true%
  ]{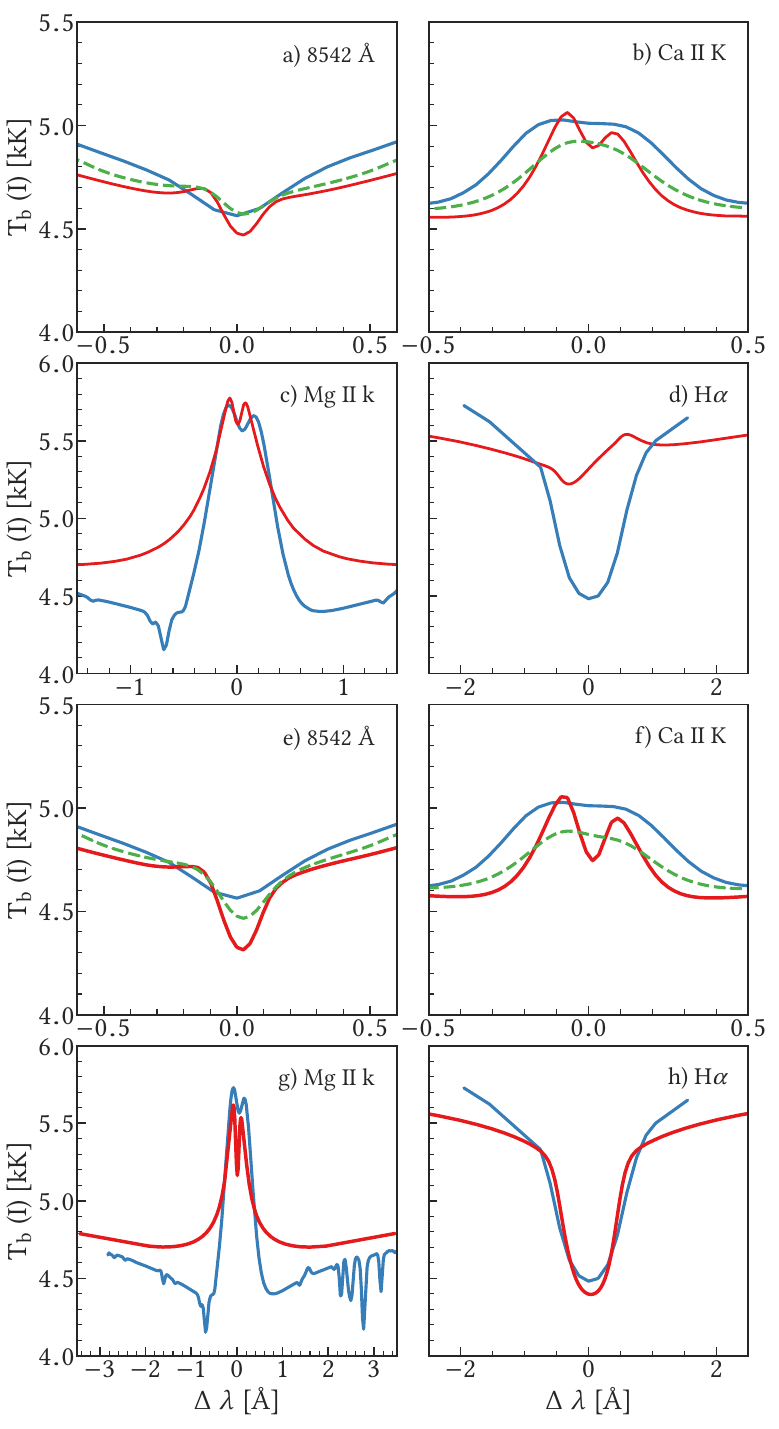}
  \caption{%
    Spatially-averaged line profiles of \CaII~8542 \AA, \CaIIK, \MgIIk, and \halpha\ expressed as brightness
    temperature $T_\text{b}$ 
    for the model atmosphere and our observations. Blue: observations; red: model atmosphere at full resolution; green dashed: model atmosphere spectrally degraded to match the resolution of the observations.
    \emph{Panels a-d}: spatial averages for the entire model field-of-view.
    \emph{Panels e-h}: spatial averages for a weakly-magnetized patch only.
    The \halpha\ and \MgIIk\ lines are not degraded spectrally.
  }
  \label{fig:average_spectrum}
\end{figure}
We compared the spatially-averaged synthetic spectra with observations
from SST and IRIS to see 
{how far the model is from the real Sun. Although the solar features represented in the observed 
and simulated samples are quite different, the comparison allows us also to place this model in the context of previous studies \citep[e.g][]{2013ApJ...772...90L,2018A&A...611A..62B}. }
Figure~\ref{fig:average_spectrum} shows the spatially-averaged profiles of
\CaIR, \CaIIK, \MgIIk, and \halpha\ from the model atmosphere and observations (see Section~\ref{sec:observation}).
Two different averages are shown for the synthetic spectra: the entire model atmosphere (panels a--d), and in a patch with relatively
weak magnetic field (panels e--h).
We note that the observations are not in any way chosen to represent all active
regions, they just illustrate
how average line profiles in active regions look like.

The observations of \halpha\ and  \CaIR\ show absorption profiles, while \MgIIk,
and \CaIIK\ show a double-peaked central reversal (marginally visible in \CaIIK\ owing
to the spectral resolution of SST/CHROMIS).
The synthetic spectra for the atmosphere outside the flare ribbons show
qualitatively the same behavior, but substantial differences in a quantitative
sense.

The averaged spectra from the model atmosphere
gives a peak separation of 16~km~s$^{-1}$ (0.15~\AA) for \MgIIk\ and
10.5~km~s$^{-1}$ (0.14~\AA) for \CaIIK. These numbers are $\sim$30\% larger than those reported for
Bifrost simulations of the quiet Sun
\citep{2013ApJ...772...90L,2018A&A...611A..62B},
but still $\sim$1.5--2 times smaller than
 observed for
the quiet Sun.

The averaged profiles of \CaIIK\ and 8\,542~\AA\ show that
the model atmosphere is colder in the lower chromosphere and photosphere than the observations.
The synthetic line wings of \MgIIk\ have a higher intensity
than  the observations.
We also note that the inner wings of the \CaIR, \CaIIK, and \halpha\ line lie
below the observed intensity, while the \MgIIk\ wings lie above it. 

Inspection of Fig.\ \ref{fig:average_spectrum}d shows that
the averaged \halpha\ profile has a central emission and
a weak and blue-shifted central absorption core, in stark contrast to the
observations.
Figure~\ref{fig:average_spectrum}h shows that the core
behaves as usual in the less magnetized patch of the simulation.
We show in Section~\ref{sec:flare_profiles} that the filling of the \halpha\ core
in the averaged profile is caused by a strong \halpha\ emission
in the flare ribbons.

\subsection{Spatially resolved synthetic profiles}
\label{sec:indivdual_synthetic_profiles}

In Figs.~\ref{fig:flare_profiles}--\ref{fig:fibril_profiles} we
explain the formation of line intensity profiles in three different
structures in the
simulation: flare ribbons (with and without strong velocity gradients), a fibril, and a sunspot umbra.
The top row of each figure shows the profiles for each line, while the bottom
row shows the temperature and line source functions for \halpha\ and
\CaIR\ in the left panel, the two-level photon destruction probability
in the middle panel, the height where $\tau_\nu = 1$ for each line
together with the vertical velocity in the right panel.
The photon destruction probability is defined as
\begin{equation}
  \epsilon = 
    \dfrac{ C_{\!ji} }{ C_{\!ji} + A_{\!ji} + B_{\!ji} B_\nu },
\end{equation}
with the downward collisional rate $C_{\!ji}$, the Einstein coefficients $A_{\!ji}$ for spontaneous deexcitation, $B_{\!ji}$ for induced deexitation, and the Planck function $B_\nu$.
We do not display the line source functions for \CaIIK\ and \MgIIk\ because they are frequency-dependent owing to PRD effects.
We show $\epsilon$ because it gives an indication of the sensitivity of the line source function to the local temperature using the two-level source function approximation: $S_\nu = (1-\epsilon) \bar{J}_{\nu0} + \epsilon B_\nu$.

\subsubsection{Flare ribbons} \label{sec:flare_profiles}
%
\begin{figure*}
  \includegraphics[%
    width = \textwidth,%
    trim = 2pt 2pt 2pt 2pt,%
    clip = true%
  ]{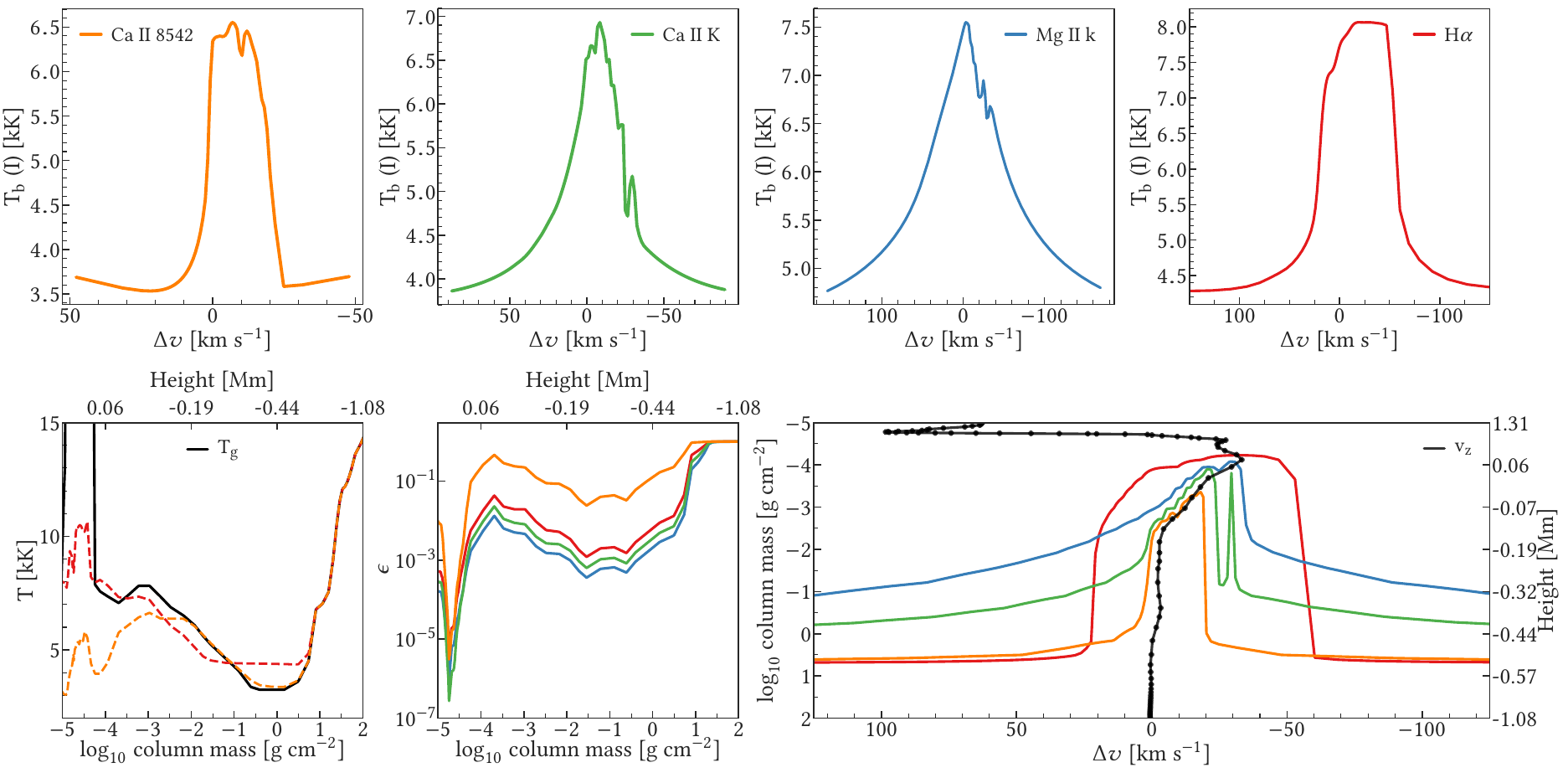}%
  \caption{%
    Line formation in a flare ribbon (orange cross in Fig.\ \ref{fig:formation_height2}).
    \emph{Top row}: profiles of vertically emergent intensity, expressed as
    brightness temperature $T_\text{b}$, as function of frequency from line center (in Doppler shift units) for \ion{Ca}{II} 8\,542~\AA\ (orange), \CaIIK\ (green),
    \MgIIk\ (blue), and \halpha\ (red).
    \emph{Bottom left}: the line source functions for H$\alpha$ (red dashed) and
    \ion{Ca}{II} 8\,542\ \AA\ (orange dashed), expressed as excitation
    temperature, and the gas temperature (solid black) as a function of column mass.
    \emph{Bottom middle}: photon destruction probabilities for each of the lines.
    \emph{Bottom right}: 
    height of optical depth unity 
    (solid colored) and the vertical velocity (solid black, dots mark the
    grid points, upflow is positive) as a function of frequency from line center in Doppler shift units.
  }
  \label{fig:flare_profiles}
\end{figure*}
\begin{figure*}
  \includegraphics[%
    width = \textwidth,%
    trim = 2pt 2pt 2pt 2pt,%
    clip = true%
  ]
  {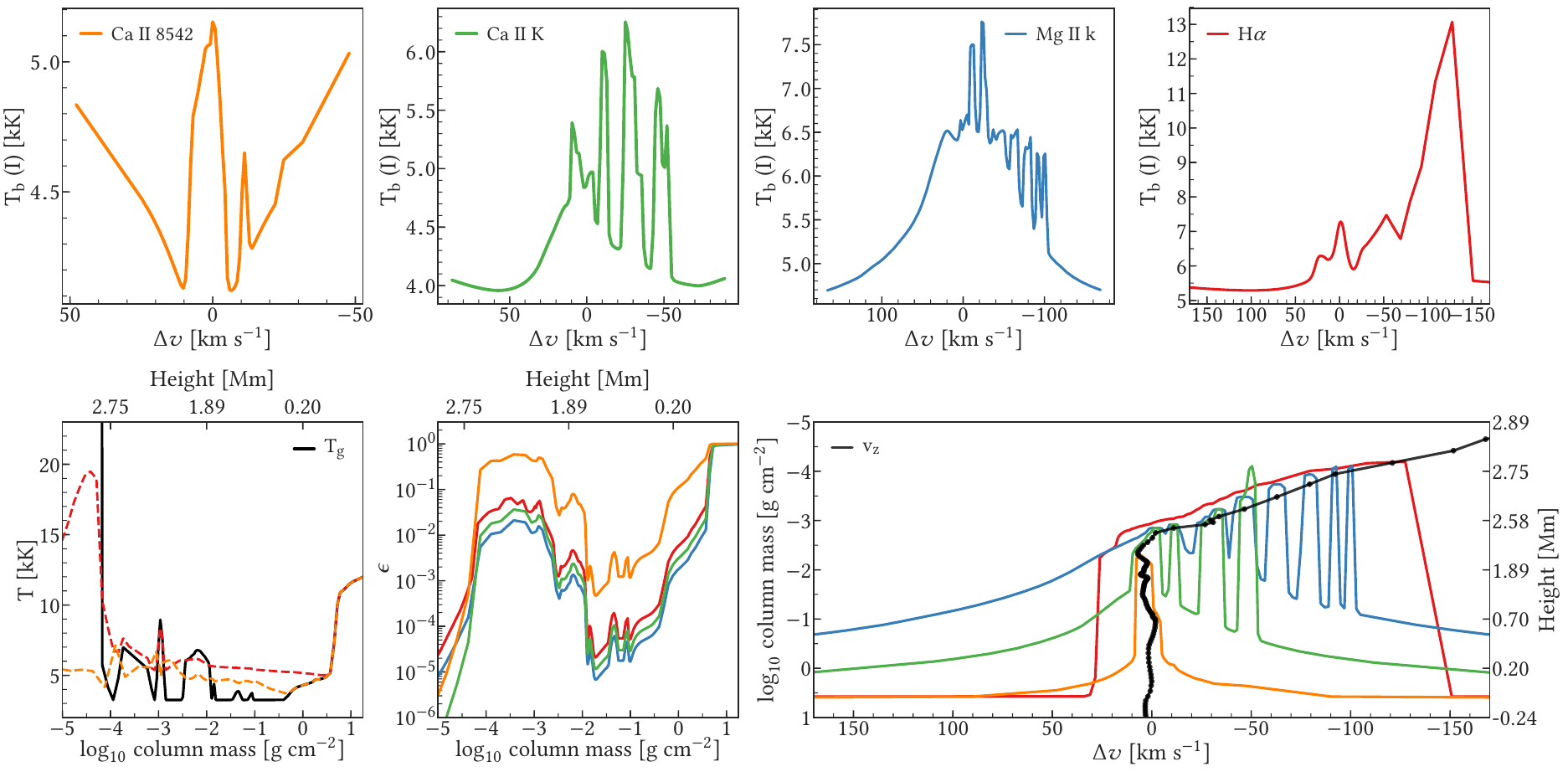}
  \caption{%
    Line formation in a flare ribbon with strong velocity gradient (blue cross
    in Fig.\ \ref{fig:formation_height2}).
    The figure follows the same format as Fig.~\ref{fig:flare_profiles}.
  }
  \label{fig:flare_profiles_2}
\end{figure*}

We investigated our model atmosphere and line formation in the flare ribbons, and show two columns in Figs.~\ref{fig:flare_profiles}--\ref{fig:flare_profiles_2}. The upper panels of Fig.~\ref{fig:flare_profiles} show  an example of broad, slightly redshifted emission peaks in all four  lines. The chromosphere has a high mass density as evidenced by the large values of $\epsilon$ (which is roughly proportional to the electron density). The transition region lies at a high column mass of $10^{-4.2}$~g~cm$^{-2}$, and the chromospheric temperature rise starts deep in the atmosphere at 0.2~g~cm$^{-2}$. This partly explains why the ribbons look so similar when the intensities are computed in 1D and 3D (see Fig.~\ref{fig:halpha_1d_vs_3d}). The strong chromospheric temperature rise leads to a source function increasing with height for all lines, leading to the emission profiles, while the deep location of the onset of the rise causes the wide symmetric base of the emission peaks, especially for \MgIIk. The downflow just below the transition region causes the redshift of the emission peak.

 These relatively smooth flare ribbon profiles are actually rather rare in our FOV. We often find more complex spectral profiles, such as shown in Fig.~\ref{fig:flare_profiles_2}.
They typically show multiple-peaked profiles in
\CaIIK\ and \MgIIk, with strongly Doppler-shifted emission peaks. The \halpha\ profile is very broad and has its strongest emission peak blueshifted by 130~km~s$^{-1}$. The temperature in the chromosphere shows multiple peaks, with low temperatures between the peaks. The mass density, and thus the photon destruction probability are again high  and the transition region is again located at a high column mass. The atmosphere differs from the one shown in Fig.~\ref{fig:flare_profiles} in that the transition region harbors a strong velocity gradient. The cores of \MgIIk\ and \CaIIK\ form completely in the chromosphere in this gradient. The strongest emission peak in \halpha\ reaches optical depth unity between the last chromospheric grid point with a temperature of 5.5~kK, and the first transition-region/coronal grid point which has a temperature of 450~kK (i.e., in the temperature gradient indicated by the nearly vertical black line in the lower-left panel of Fig.~\ref{fig:flare_profiles_2}.

The finite vertical resolution of the model atmosphere produces the peculiar, multiple-peaked line shapes of \MgIIk, and \CaIIK. The difference in vertical velocity between adjacent grid points is larger than the Doppler width, and this causes spurious emission peaks, as explained by
\citet{2013A&A...549A.126I}.
This is reflected in the peaks in the $z(\tau_\nu = 1)$ curves in the lower right panel.
The formation height is large at the velocities of the grid points (black filled circles), but drops to low values when more than a Doppler width away from the velocities at the grid points.

{The multiple emission peaks are thus artefact produced by the limited spatial resolution in the radiative transfer computation. Had the velocity gradient been properly resolved, then these line profiles would show a broad asymmetric emission peak.}

The \halpha\ profile does not show such strong emission peaks.
The emission peak at $\Delta\varv = -120$~km~s$^{-1}$ is, however, badly resolved. The frequency grid is not fine enough to resolve the right flank, and as stated before, the peak forms within a single grid interval, which constitutes the interface between the chromosphere and transition region (compare the brightness temperature of the emission peak of 13 kK to the source function in the lower right panel). In this interval the optical depth increases from $10^{-5}$ to 1.2. The contribution function is thus not properly resolved, and the resulting intensity depends on the chosen solution scheme. 
We investigated the source function, and exactly in this grid interval its changes from two-level behavior in the chromosphere to recombination-radiation-dominated in the transition region.

\subsubsection{Sunspot}
%
\begin{figure*}
  \includegraphics[%
    width = \textwidth,%
    trim = 2pt 2pt 2pt 2pt,%
    clip = true%
  ]{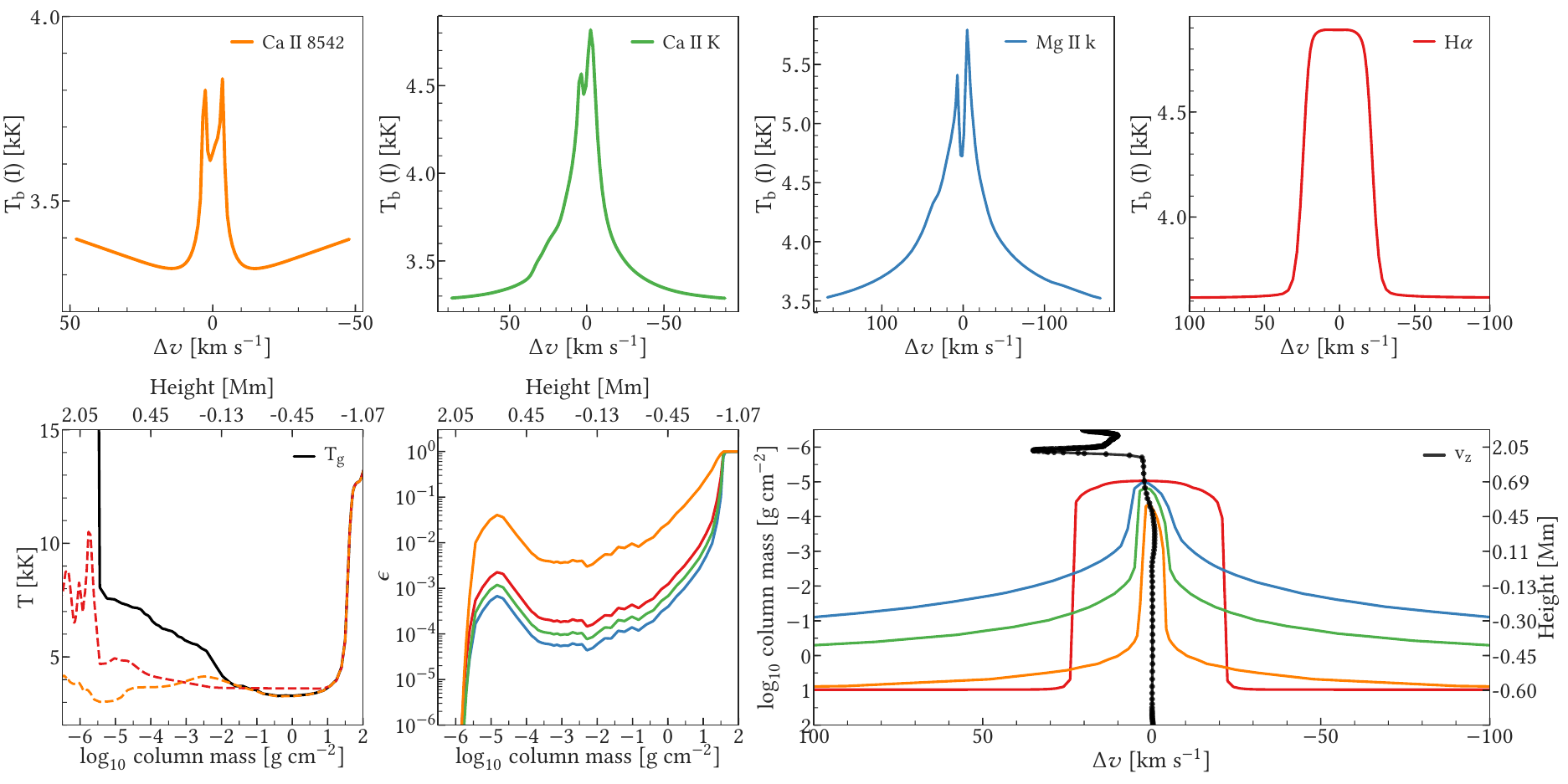}%
  \caption{%
    Line formation in a sunspot (green cross in Fig.\ \ref{fig:formation_height2}).
    The figure follows the same format as Fig.~\ref{fig:flare_profiles}.
  }
  \label{fig:sunspot_profiles}
\end{figure*}
Figure \ref{fig:sunspot_profiles} shows the line formation in a sunspot in the model atmosphere. The \halpha\ core is in emission, and the three other spectral lines show double-peaked profiles. This is common in the model atmosphere, we find only a few locations where \halpha\ is in absorption and the other lines have single emission peaks. 

The temperature stratification as shown in Fig.~\ref{fig:sunspot_profiles} is common across the sunspots, from the temperature minimum the gas temperature monotonically increases towards the transition region. The \halpha\ line is in emission because the line core is formed in the steep temperature gradient of the transition region, with a high source function. The line cores of the other lines tend to be formed just below the transition region. The chromosphere has a temperature rise that is deep enough to cause emission peaks.  

Comparing to many available observations
\citep[e.g.][]{2018ApJ...861...62P,2013SoPh..288...73M,2013A&A...556A.115D,2011JPhCS.271a2040F,1967SoPh....2..234E},
we see that the simulated sunspot profiles are somewhat unusual: observed \halpha\ profiles are typically in absorption; the \CaIIK\ and \MgIIk\ profiles are typically, but not always, single peaked; \CaIR\ is in absorption and only shows emission when an umbral flash passes through.

\subsubsection{Fibril}
\label{sec:fibril}

\begin{figure*}
  \includegraphics[%
    width = \textwidth,%
    trim = 2pt 2pt 2pt 2pt,%
    clip = true%
  ]{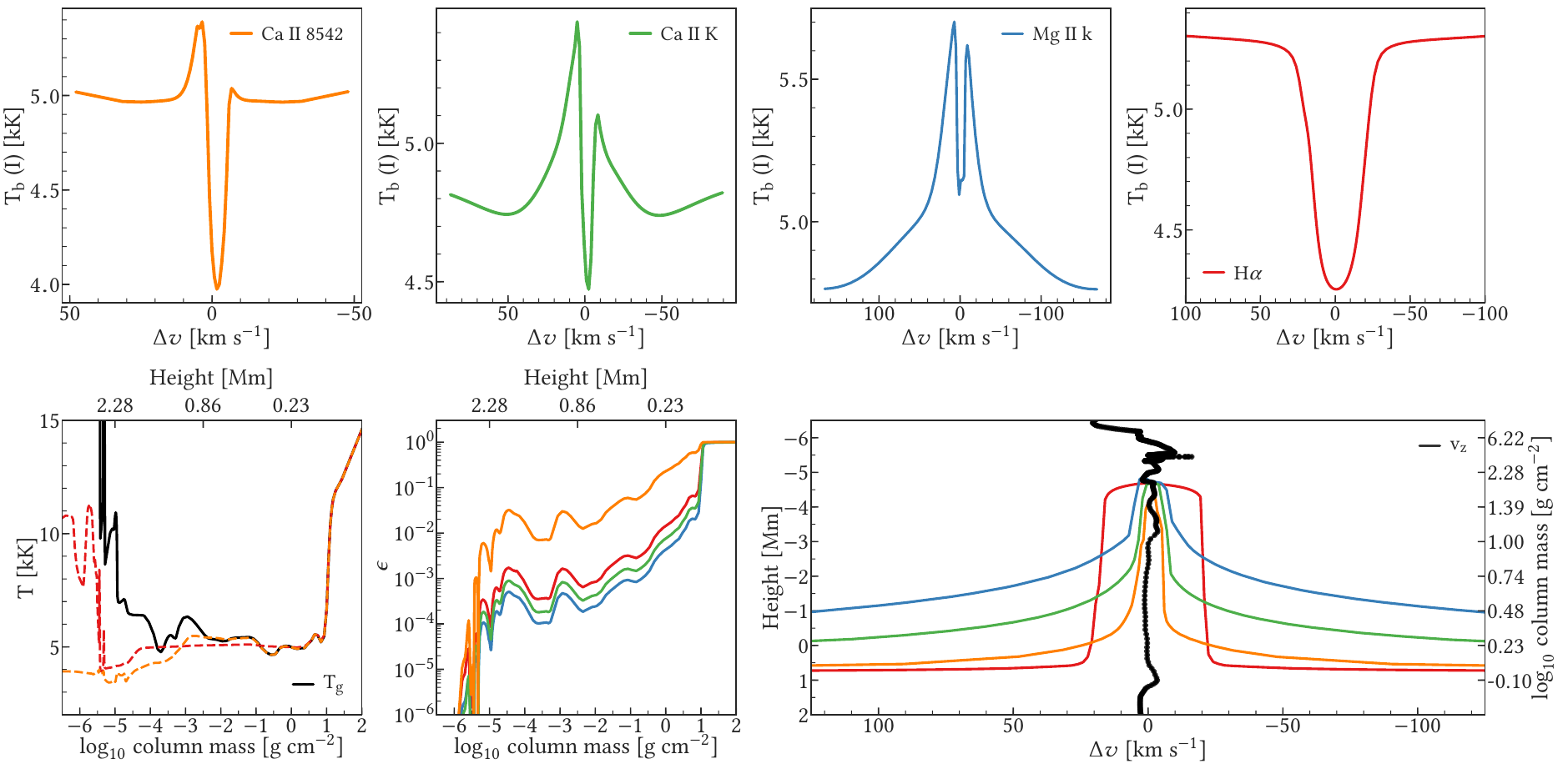}%
  \caption{%
    Line formation in a fibril (red cross in Fig.\ \ref{fig:formation_height2}). The figure follows the same format as Fig.~\ref{fig:flare_profiles}.
  }
  \label{fig:fibril_profiles}
\end{figure*}

Figure~\ref{fig:fibril_profiles} shows the typical line formation in a fibril.
Here \halpha\ is in absorption, \CaIIK, \MgIIk, and \CaIR\ each have an
asymmetric double-peaked emission core.
We find that this situation is common for \halpha, \CaIIK, and \MgIIk. The density in the chromosphere is relatively low, and $\epsilon$ is an order of magnitude smaller than in the flare-ribbon columns.
The \halpha\ line core is formed below the transition region
and therefore is not influenced by the increase in the
source function there.

The \CaIR\ line shows much more variation, ranging from pure absorption to wide and asymmetric profiles with three emission peaks. The line is mainly in absorption in fibrils in our data, while we also find some instances of profiles with a double-peaked central reversal as in Fig.~\ref{fig:fibril_profiles}.

\section{Discussion and conclusions} \label{sec:discussion}

%
\citet{2018NatAs.tmp..173C} 
performed a radiation-MHD simulation of an active region including emergence of
a new flux that produced a flare-like reconnection event in the corona. We took a snapshot from a rerun of their simulation with twice the original numerical resolution.

This numerical experiment, aimed only at investigating the corona, produced a
unique data set suitable for investigating the chromosphere as well.
The simulation domain is geometrically $4\times 2\times 3$ times bigger and models a much more active chromosphere
than the commonly-used publicly-available Bifrost simulation
\citep[][]{2016A&A...585A...4C}, 

The trade-off for getting the large domain size is that the grid
spacing, 96~km horizontally and 32~km vertically, is relatively coarse.
Radiative energy losses in the photosphere and corona are approximated using
the gray LTE transfer.
The absorption of coronal radiation in the chromosphere is not included.
The equation of state assumes LTE, whereas hydrogen and helium both require
non-equilibrium ionization
\citep{2007A&A...473..625L,2014ApJ...784...30G}.
There are no interactions between ions and neutral species, which significantly
structure and heat the chromosphere
\citep{2012ApJ...753..161M,Martinez-Sykora1269}.
{These simplifications have an impact on the structure of the chromosphere, both for the temperature and density in the model atmosphere.}
However, these simplifications do not affect the large-scale magnetic structure in
the chromosphere.
Neither do they distort the formation of flare ribbons, chromospheric evaporations
through heat conduction, even though the exact location of the transition region
changes if more accurate radiative losses are included.

In this paper we have taken one snapshot of this MURaM simulation and have
solved numerically the problem of 3D radiative transfer for \HI, \CaII, and
\MgII\ in non-LTE including partially-coherent line scattering (PRD) in the
\MgIIk\ and \CaIIK\ lines.
We have corrected the electron density by treating hydrogen in
statistical equilibrium together with an equation for charge conservations.
On the radiative transfer side, this is a large improvement over the electron
density evaluated in LTE.
For the sake of numerical stability and computational speed, we have set the
minimum temperature in the model atmosphere to 3\,250~K, we have used a simple
approximation for the pressure broadening of \halpha, and we have limited the
size of our \HI\ model atom.

The finite grid spacing resulted in a poor sampling in both
frequency and optical depth that led to artefacts in the line profiles when the
line-forming region contains large gradients in
temperature, velocity and/or density.
While these minor weaknesses should be addressed in future work to allow
detailed quantitative comparison with observations, they have no consequences
for the qualitative results shown this paper.

We have shown that 3D non-LTE radiative transfer
computations including charge conservation and PRD
can handle a large variation in all physical quantities, in a different
parameter regime (active region) than studied before
\citep[more quiet-sun-like,][]{2016A&A...585A...4C}. 

The synthetic line-core images have shown long
($\sim$35~Mm) strands connecting the opposite-polarity sunspots, that resemble fibrils seen in
observations. We have established that the strong fibrils seen in
\halpha\ are mostly aligned with the magnetic field direction.
Fibrils seen in the other lines are less aligned, especially those in
\ion{Ca}{II}~8\,542~\AA, which suffer from sudden changes of the height of optical
depth unity so.
This contrasts with the findings in 
\citet{2015ApJ...802..136L}, 
who found that in more quiet-Sun circumstances
only a fraction of the fibrils seen in \halpha\ follow the same field line.

In 3D MHD simulations of quiet Sun,
\citet{2012ApJ...749..136L,2013ApJ...772...90L} 
found that the cores of \ion{Mg}{II}~k and \ion{Ca}{II} are both formed in the upper
chromosphere in a wide range of heights following the transition region while
the cores of H$\alpha$ and \ion{Ca}{II} 8\,542~\AA\ are formed at lower heights, in the middle chromosphere.

We have found two important differences with this in the MURaM simulation.
Firstly, the H$\alpha$ line is formed in a much larger range of heights.
Secondly, the cores of H$\alpha$, \ion{Mg}{II}~k, and \ion{Ca}{II}~K are formed
at very similar heights while the core of \ion{Ca}{II} 8\,542~\AA\ is formed
only some 150~km below them. {This qualitatively agrees with observations of active regions, which show a very similar appearance in the cores of  \halpha\ and \CaIIK\
\citet{2018A&A...612A..28L}.
Given that the model used in this paper manages to reproduce overal apparence in all chromospheric observables we synhesized, indicates that the model gives a decent representation of long solar fibrils.}

The flare ribbons in this simulation are caused by thermal conduction only.
They appear bright in the line cores because the thermal conduction and dense coronal loops above the ribbon cause
a hot and dense upper chromosphere, with sufficient electron density
to couple the source function to the local temperature.
This mechanism is similar to what
\citet{2015ApJ...809L..30C} 
proposed for \MgIIk\ in plages. {We find synthetic line profiles that are broad, asymmetric, and with a single emission peak, similar to the profiles observed in flare ribbons.} 

The flare ribbons are poorly visible in \CaIR, in contrast to observations
\citep[e.g.,][]{2012ApJ...748..138K}. 
This agrees with the standard flare model that
requires beams of non-thermal electrons
or Alfv\'enic waves to deposit thermal energy in the middle chromosphere.
Nevertheless, the visibility of the flare ribbons
hints that the fraction of the flare
energy transported down by thermal conduction is enough to significantly alter
the structure of the chromosphere.

This study is just a first step to understand the chromosphere in active regions
using 3D radiation-MHD simulations.
In our study, we have only touched on the formation of the
strongest chromospheric lines.
Future studies should look into the formation of the lines in more detail.

\begin{acknowledgements}
The Swedish 1-m Solar Telescope is operated on the island of La Palma by the
Institute for Solar Physics of Stockholm University in the Spanish Observatorio
del Roque de los Muchachos of the Instituto de Astrofísica de Canarias.
The computations were performed on resources provided by the Swedish National
Infrastructure for Computing (SNIC) at the High Performance Computing Center
North at Ume\aa\ University and the PDC Centre for High Performance Computing
(PDC-HPC) at the Royal Institute of Technology in Stockholm.
JL was supported by a grant (2016.0019) of the Knut och Alice Wallenberg
foundation.
SD and JdlCR were supported by a grant from the Swedish Civil Contingencies
Agency (MSB).
This study has been discussed within the activities of team 399 `Studying
magnetic-field-regulated heating in the solar chromosphere' at the International
Space Science Institute (ISSI) in Switzerland.
JdlCR is supported by grants from the Swedish Research Council (2015-03994), the
Swedish National Space Board (128/15).
This project has received funding from the European Research Council (ERC) under
the European Union's Horizon 2020 research and innovation programme (SUNMAG,
grant agreement 759548).
AVS acknowledges financial support from the Spanish Ministry of Economy and
Competitiveness (MINECO) under the 2015 Severo Ochoa Program MINECO
SEV-2015-0548.
The National Center for Atmospheric Research is sponsored by the National Science Foundation. 
We would like to thank Adam F.\ Kowalski and Joel C.\ Allred for providing us
their modified RH code, which includes the unified theory of electric pressure
broadening.
JPB would thank Gregal J.\ M.\ Vissers for the help in calibrating the IRIS
data, and Carolina Robustini for discussions.
\end{acknowledgements}

\bibliographystyle{aa} 
\bibliography{article}

\begin{thebibliography}{78}
\expandafter\ifx\csname natexlab\endcsname\relax\def\natexlab#1{#1}\fi

\bibitem[{{Allred} {et~al.}(2005){Allred}, {Hawley}, {Abbett}, \&
  {Carlsson}}]{2005ApJ...630..573A}
{Allred}, J.~C., {Hawley}, S.~L., {Abbett}, W.~P., \& {Carlsson}, M. 2005,
  \apj, 630, 573

\bibitem[{{Arber} {et~al.}(2007){Arber}, {Haynes}, \&
  {Leake}}]{2007ApJ...666..541A}
{Arber}, T.~D., {Haynes}, M., \& {Leake}, J.~E. 2007, \apj, 666, 541

\bibitem[{{Aschwanden} {et~al.}(2016){Aschwanden}, {Reardon}, \&
  {Jess}}]{2016ApJ...826...61A}
{Aschwanden}, M.~J., {Reardon}, K., \& {Jess}, D.~B. 2016, \apj, 826, 61

\bibitem[{{Asensio Ramos} {et~al.}(2017){Asensio Ramos}, {de la Cruz
  Rodr{\'\i}guez}, {Mart{\'\i}nez Gonz{\'a}lez}, \&
  {Socas-Navarro}}]{2017A&A...599A.133A}
{Asensio Ramos}, A., {de la Cruz Rodr{\'\i}guez}, J., {Mart{\'\i}nez
  Gonz{\'a}lez}, M.~J., \& {Socas-Navarro}, H. 2017, \aap, 599

\bibitem[{{Beckers}(1964)}]{1964PhDT........83B}
{Beckers}, J.~M. 1964, PhD thesis, Sacramento Peak Observatory, Air Force
  Cambridge Research Laboratories, Mass., USA

\bibitem[{{Benz}(2008)}]{2008LRSP....5....1B}
{Benz}, A.~O. 2008, Living Reviews in Solar Physics, 5, 1

\bibitem[{{Bj{\o}rgen} {et~al.}(2018){Bj{\o}rgen}, {Sukhorukov}, {Leenaarts},
  {Carlsson}, {de la Cruz Rodr{\'\i}guez}, {Scharmer}, \&
  {Hansteen}}]{2018A&A...611A..62B}
{Bj{\o}rgen}, J.~P., {Sukhorukov}, A.~V., {Leenaarts}, J., {et~al.} 2018, \aap,
  611

\bibitem[{{Carlson}(1963)}]{carlson1963}
{Carlson}, B.~G. 1963, in Methods in Computational Physics, Vol. 1, Statistical
  Physics, Methods in Computational Physics: Advances in Research and
  Applications, ed. B.~{Alder}, S.~{Fernbach}, \& M.~{Rotenberg} (New York, NY,
  USA: Academic Press), 1--42, 304~p.

\bibitem[{{Carlsson} {et~al.}(2016){Carlsson}, {Hansteen}, {Gudiksen},
  {Leenaarts}, \& {De Pontieu}}]{2016A&A...585A...4C}
{Carlsson}, M., {Hansteen}, V.~H., {Gudiksen}, B.~V., {Leenaarts}, J., \& {De
  Pontieu}, B. 2016, \aap, 585, A4

\bibitem[{{Carlsson} \& {Leenaarts}(2012)}]{2012A&A...539A..39C}
{Carlsson}, M. \& {Leenaarts}, J. 2012, \aap, 539, A39

\bibitem[{{Carlsson} {et~al.}(2015){Carlsson}, {Leenaarts}, \& {De
  Pontieu}}]{2015ApJ...809L..30C}
{Carlsson}, M., {Leenaarts}, J., \& {De Pontieu}, B. 2015, \apjl, 809, L30

\bibitem[{{Carlsson} \& {Stein}(1992)}]{1992ApJ...397L..59C}
{Carlsson}, M. \& {Stein}, R.~F. 1992, \apjl, 397, L59

\bibitem[{{Carlsson} \& {Stein}(2002)}]{2002ApJ...572..626C}
{Carlsson}, M. \& {Stein}, R.~F. 2002, \apj, 572, 626

\bibitem[{{Cheung} \& {Cameron}(2012)}]{2012ApJ...750....6C}
{Cheung}, M.~C.~M. \& {Cameron}, R.~H. 2012, \apj, 750, 6

\bibitem[{{Cheung} \& {Isobe}(2014)}]{2014LRSP...11....3C}
{Cheung}, M. C.~M. \& {Isobe}, H. 2014, Living Reviews in Solar Physics, 11, 3

\bibitem[{{Cheung} {et~al.}(2018){Cheung}, {Rempel}, {Chintzoglou}, {Chen},
  {Testa}, {Mart{\'\i}nez-Sykora}, {Sainz Dalda}, {DeRosa}, {Malanushenko},
  {Hansteen}, {De Pontieu}, {Carlsson}, {Gudiksen}, \&
  {McIntosh}}]{2018NatAs.tmp..173C}
{Cheung}, M.~C.~M., {Rempel}, M., {Chintzoglou}, G., {et~al.} 2018, Nature
  Astronomy, 173

\bibitem[{{Danilovic}(2017)}]{2017A&A...601A.122D}
{Danilovic}, S. 2017, \aap, 601, A122

\bibitem[{{de la Cruz Rodr{\'\i}guez} {et~al.}(2018){de la Cruz
  Rodr{\'\i}guez}, {Leenaarts}, {Danilovic}, \&
  {Uitenbroek}}]{2018arXiv181008441D}
{de la Cruz Rodr{\'\i}guez}, J., {Leenaarts}, J., {Danilovic}, S., \&
  {Uitenbroek}, H. 2018, arXiv e-prints, arXiv:1810.08441

\bibitem[{{de la Cruz Rodr{\'\i}guez} {et~al.}(2015){de la Cruz
  Rodr{\'\i}guez}, {L{\"o}fdahl}, {S{\"u}tterlin}, {Hillberg}, \& {Rouppe van
  der Voort}}]{2015A&A...573A..40D}
{de la Cruz Rodr{\'\i}guez}, J., {L{\"o}fdahl}, M.~G., {S{\"u}tterlin}, P.,
  {Hillberg}, T., \& {Rouppe van der Voort}, L. 2015, \aap, 573, A40

\bibitem[{{de la Cruz Rodr{\'{\i}}guez} {et~al.}(2013){de la Cruz
  Rodr{\'{\i}}guez}, {Rouppe van der Voort}, {Socas-Navarro}, \& {van
  Noort}}]{2013A&A...556A.115D}
{de la Cruz Rodr{\'{\i}}guez}, J., {Rouppe van der Voort}, L., {Socas-Navarro},
  H., \& {van Noort}, M. 2013, \aap, 556, A115

\bibitem[{{de la Cruz Rodr{\'\i}guez} \&
  {Socas-Navarro}(2011)}]{2011A&A...527L...8D}
{de la Cruz Rodr{\'\i}guez}, J. \& {Socas-Navarro}, H. 2011, \aap, 527

\bibitem[{{De Pontieu} {et~al.}(2014){De Pontieu}, {Title}, {Lemen}, {Kushner},
  {Akin}, {Allard}, {Berger}, {Boerner}, {Cheung}, {Chou}, {Drake}, {Duncan},
  {Freeland}, {Heyman}, {Hoffman}, {Hurlburt}, {Lindgren}, {Mathur}, {Rehse},
  {Sabolish}, {Seguin}, {Schrijver}, {Tarbell}, {W{\"u}lser}, {Wolfson},
  {Yanari}, {Mudge}, {Nguyen-Phuc}, {Timmons}, {van Bezooijen}, {Weingrod},
  {Brookner}, {Butcher}, {Dougherty}, {Eder}, {Knagenhjelm}, {Larsen},
  {Mansir}, {Phan}, {Boyle}, {Cheimets}, {DeLuca}, {Golub}, {Gates}, {Hertz},
  {McKillop}, {Park}, {Perry}, {Podgorski}, {Reeves}, {Saar}, {Testa}, {Tian},
  {Weber}, {Dunn}, {Eccles}, {Jaeggli}, {Kankelborg}, {Mashburn}, {Pust},
  {Springer}, {Carvalho}, {Kleint}, {Marmie}, {Mazmanian}, {Pereira}, {Sawyer},
  {Strong}, {Worden}, {Carlsson}, {Hansteen}, {Leenaarts}, {Wiesmann},
  {Aloise}, {Chu}, {Bush}, {Scherrer}, {Brekke}, {Martinez-Sykora}, {Lites},
  {McIntosh}, {Uitenbroek}, {Okamoto}, {Gummin}, {Auker}, {Jerram}, {Pool}, \&
  {Waltham}}]{2014SoPh..289.2733D}
{De Pontieu}, B., {Title}, A.~M., {Lemen}, J.~R., {et~al.} 2014, \solphys, 289,
  2733

\bibitem[{{Engvold}(1967)}]{1967SoPh....2..234E}
{Engvold}, O. 1967, \solphys, 2, 234

\bibitem[{{Felipe} {et~al.}(2011){Felipe}, {Khomenko}, {Collados}, \&
  {Beck}}]{2011JPhCS.271a2040F}
{Felipe}, T., {Khomenko}, E., {Collados}, M., \& {Beck}, C. 2011, GONG-SoHO 24:
  A New Era of Seismology of the Sun and Solar-Like Stars, 271, 012040

\bibitem[{{Fletcher} {et~al.}(2011){Fletcher}, {Dennis}, {Hudson}, {Krucker},
  {Phillips}, {Veronig}, {Battaglia}, {Bone}, {Caspi}, {Chen}, {Gallagher},
  {Grigis}, {Ji}, {Liu}, {Milligan}, \& {Temmer}}]{2011SSRv..159...19F}
{Fletcher}, L., {Dennis}, B.~R., {Hudson}, H.~S., {et~al.} 2011, \ssr, 159, 19

\bibitem[{{Fontenla} {et~al.}(1993){Fontenla}, {Avrett}, \&
  {Loeser}}]{1993ApJ...406..319F}
{Fontenla}, J.~M., {Avrett}, E.~H., \& {Loeser}, R. 1993, \apj, 406, 319

\bibitem[{{Golding} {et~al.}(2014){Golding}, {Carlsson}, \&
  {Leenaarts}}]{2014ApJ...784...30G}
{Golding}, T.~P., {Carlsson}, M., \& {Leenaarts}, J. 2014, \apj, 784, 30

\bibitem[{{Golding} {et~al.}(2016){Golding}, {Leenaarts}, \&
  {Carlsson}}]{2016ApJ...817..125G}
{Golding}, T.~P., {Leenaarts}, J., \& {Carlsson}, M. 2016, \apj, 817, 125

\bibitem[{{Gudiksen} {et~al.}(2011){Gudiksen}, {Carlsson}, {Hansteen}, {Hayek},
  {Leenaarts}, \& {Mart{\'{\i}}nez-Sykora}}]{2011A&A...531A.154G}
{Gudiksen}, B.~V., {Carlsson}, M., {Hansteen}, V.~H., {et~al.} 2011, \aap, 531,
  A154

\bibitem[{{Hansteen} {et~al.}(2017){Hansteen}, {Archontis}, {Pereira},
  {Carlsson}, {Rouppe van der Voort}, \& {Leenaarts}}]{2017ApJ...839...22H}
{Hansteen}, V.~H., {Archontis}, V., {Pereira}, T.~M.~D., {et~al.} 2017, \apj,
  839

\bibitem[{{Hansteen} {et~al.}(2006){Hansteen}, {De Pontieu}, {Rouppe van der
  Voort}, {van Noort}, \& {Carlsson}}]{2006ApJ...647L..73H}
{Hansteen}, V.~H., {De Pontieu}, B., {Rouppe van der Voort}, L., {van Noort},
  M., \& {Carlsson}, M. 2006, \apjl, 647, L73

\bibitem[{{Heinzel}(1995)}]{1995A&A...299..563H}
{Heinzel}, P. 1995, \aap, 299, 563

\bibitem[{{Ibgui} {et~al.}(2013){Ibgui}, {Huben{\'{y}}}, {Lanz}, \&
  {Stehl{\'e}}}]{2013A&A...549A.126I}
{Ibgui}, L., {Huben{\'{y}}}, I., {Lanz}, T., \& {Stehl{\'e}}, C. 2013, \aap,
  549, A126

\bibitem[{{Judge} {et~al.}(2011){Judge}, {Tritschler}, \& {Chye
  Low}}]{2011ApJ...730L...4J}
{Judge}, P.~G., {Tritschler}, A., \& {Chye Low}, B. 2011, \apjl, 730, L4

\bibitem[{{Kerr} {et~al.}(2015){Kerr}, {Sim{\~o}es}, {Qiu}, \&
  {Fletcher}}]{2015A&A...582A..50K}
{Kerr}, G.~S., {Sim{\~o}es}, P.~J.~A., {Qiu}, J., \& {Fletcher}, L. 2015, \aap,
  582

\bibitem[{{Kleint}(2012)}]{2012ApJ...748..138K}
{Kleint}, L. 2012, \apj, 748, 138

\bibitem[{{Koesterke} {et~al.}(2002){Koesterke}, {Hamann}, \&
  {Gr{\"a}fener}}]{2002A&A...384..562K}
{Koesterke}, L., {Hamann}, W.-R., \& {Gr{\"a}fener}, G. 2002, \aap, 384, 562

\bibitem[{{Kowalski} {et~al.}(2017){Kowalski}, {Allred}, {Uitenbroek},
  {Tremblay}, {Brown}, {Carlsson}, {Osten}, {Wisniewski}, \&
  {Hawley}}]{2017ApJ...837..125K}
{Kowalski}, A.~F., {Allred}, J.~C., {Uitenbroek}, H., {et~al.} 2017, \apj, 837,
  125

\bibitem[{{Kuridze} {et~al.}(2012){Kuridze}, {Morton}, {Erd{\'e}lyi},
  {Dorrian}, {Mathioudakis}, {Jess}, \& {Keenan}}]{2012ApJ...750...51K}
{Kuridze}, D., {Morton}, R.~J., {Erd{\'e}lyi}, R., {et~al.} 2012, \apj, 750, 51

\bibitem[{{Leenaarts} \& {Carlsson}(2009)}]{2009ASPC..415...87L}
{Leenaarts}, J. \& {Carlsson}, M. 2009, in ASP Conference Series, Vol. 415, The
  Second Hinode Science Meeting: Beyond Discovery---Toward Understanding, ed.
  B.~{Lites}, M.~{Cheung}, T.~{Magara}, J.~{Mariska}, \& K.~{Reeves},
  Astronomical Society of the Pacific (390 Ashton Avenue, San Francisco, CA,
  USA: Sheridan Books, Ann Arbor, MI, USA), 87--90

\bibitem[{{Leenaarts} {et~al.}(2007){Leenaarts}, {Carlsson}, {Hansteen}, \&
  {Rutten}}]{2007A&A...473..625L}
{Leenaarts}, J., {Carlsson}, M., {Hansteen}, V., \& {Rutten}, R.~J. 2007, \aap,
  473, 625

\bibitem[{{Leenaarts} {et~al.}(2012){Leenaarts}, {Carlsson}, \& {Rouppe van der
  Voort}}]{2012ApJ...749..136L}
{Leenaarts}, J., {Carlsson}, M., \& {Rouppe van der Voort}, L. 2012, \apj, 749,
  136

\bibitem[{{Leenaarts} {et~al.}(2015){Leenaarts}, {Carlsson}, \& {Rouppe van der
  Voort}}]{2015ApJ...802..136L}
{Leenaarts}, J., {Carlsson}, M., \& {Rouppe van der Voort}, L. 2015, \apj, 802,
  136

\bibitem[{{Leenaarts} {et~al.}(2018){Leenaarts}, {de la Cruz Rodr{\'\i}guez},
  {Danilovic}, {Scharmer}, \& {Carlsson}}]{2018A&A...612A..28L}
{Leenaarts}, J., {de la Cruz Rodr{\'\i}guez}, J., {Danilovic}, S., {Scharmer},
  G., \& {Carlsson}, M. 2018, \aap, 612

\bibitem[{{Leenaarts} {et~al.}(2013{\natexlab{a}}){Leenaarts}, {Pereira},
  {Carlsson}, {Uitenbroek}, \& {De Pontieu}}]{2013ApJ...772...89L}
{Leenaarts}, J., {Pereira}, T.~M.~D., {Carlsson}, M., {Uitenbroek}, H., \& {De
  Pontieu}, B. 2013{\natexlab{a}}, \apj, 772, 89

\bibitem[{{Leenaarts} {et~al.}(2013{\natexlab{b}}){Leenaarts}, {Pereira},
  {Carlsson}, {Uitenbroek}, \& {De Pontieu}}]{2013ApJ...772...90L}
{Leenaarts}, J., {Pereira}, T.~M.~D., {Carlsson}, M., {Uitenbroek}, H., \& {De
  Pontieu}, B. 2013{\natexlab{b}}, \apj, 772, 90

\bibitem[{{L{\"o}fdahl} {et~al.}(2018){L{\"o}fdahl}, {Hillberg}, {de la Cruz
  Rodriguez}, {Vissers}, {Scharmer}, {Hagfors Haugan}, \&
  {Fredvik}}]{2018arXiv180403030L}
{L{\"o}fdahl}, M.~G., {Hillberg}, T., {de la Cruz Rodriguez}, J., {et~al.}
  2018, arXiv e-prints, arXiv:1804.03030

\bibitem[{{Marsh}(1976)}]{1976SoPh...50...37M}
{Marsh}, K.~A. 1976, \solphys, 50, 37

\bibitem[{{Mart{\'\i}nez-Sykora} {et~al.}(2016){Mart{\'\i}nez-Sykora}, {De
  Pontieu}, {Carlsson}, \& {Hansteen}}]{2016ApJ...831L...1M}
{Mart{\'\i}nez-Sykora}, J., {De Pontieu}, B., {Carlsson}, M., \& {Hansteen}, V.
  2016, \apj, 831, L1

\bibitem[{Mart{\'\i}nez-Sykora {et~al.}(2012)Mart{\'\i}nez-Sykora, {De
  Pontieu}, \& {Hansteen}}]{2012ApJ...753..161M}
Mart{\'\i}nez-Sykora, J., {De Pontieu}, B., \& {Hansteen}, V. 2012, \apj, 753,
  161

\bibitem[{Mart{\'\i}nez-Sykora {et~al.}(2017)Mart{\'\i}nez-Sykora, De~Pontieu,
  Hansteen, Rouppe van~der Voort, Carlsson, \& Pereira}]{Martinez-Sykora1269}
Mart{\'\i}nez-Sykora, J., De~Pontieu, B., Hansteen, V.~H., {et~al.} 2017,
  Science, 356, 1269

\bibitem[{{Maurya} {et~al.}(2013){Maurya}, {Chae}, {Park}, {Yang}, {Song}, \&
  {Cho}}]{2013SoPh..288...73M}
{Maurya}, R.~A., {Chae}, J., {Park}, H., {et~al.} 2013, \solphys, 288, 73

\bibitem[{{Mili{\'c}} \& {van Noort}(2018)}]{2018A&A...617A..24M}
{Mili{\'c}}, I. \& {van Noort}, M. 2018, \aap, 617, A24

\bibitem[{{Nordlund}(1982)}]{1982A&A...107....1N}
{Nordlund}, A. 1982, \aap, 107, 1

\bibitem[{{Olson} \& {Kunasz}(1987)}]{1987JQSRT..38..325O}
{Olson}, G.~L. \& {Kunasz}, P.~B. 1987, Journal of Quantitative Spectroscopy
  and Radiative Transfer, 38, 325

\bibitem[{{Paletou}(1995)}]{1995A&A...302..587P}
{Paletou}, F. 1995, \aap, 302, 587

\bibitem[{{Panos} {et~al.}(2018){Panos}, {Kleint}, {Huwyler}, {Krucker},
  {Melchior}, {Ullmann}, \& {Voloshynovskiy}}]{2018ApJ...861...62P}
{Panos}, B., {Kleint}, L., {Huwyler}, C., {et~al.} 2018, \apj, 861, 62

\bibitem[{{Pietarila} {et~al.}(2009){Pietarila}, {Hirzberger}, {Zakharov}, \&
  {Solanki}}]{2009A&A...502..647P}
{Pietarila}, A., {Hirzberger}, J., {Zakharov}, V., \& {Solanki}, S.~K. 2009,
  \aap, 502, 647

\bibitem[{Press(2007)}]{press2007numerical}
Press, W.~H. 2007, Numerical recipes 3rd edition: The art of scientific
  computing (Cambridge university press)

\bibitem[{{Reardon} {et~al.}(2011){Reardon}, {Wang}, {Muglach}, \&
  {Warren}}]{2011ApJ...742..119R}
{Reardon}, K.~P., {Wang}, Y.-M., {Muglach}, K., \& {Warren}, H.~P. 2011, \apj,
  742, 119

\bibitem[{{Rempel}(2017)}]{2017ApJ...834...10R}
{Rempel}, M. 2017, \apj, 834, 10

\bibitem[{{Robustini} {et~al.}(2019){Robustini}, {Esteban Pozuelo},
  {Leenaarts}, \& {de la Cruz Rodr{\'{\i}}guez}}]{2019A&A...621A...1R}
{Robustini}, C., {Esteban Pozuelo}, S., {Leenaarts}, J., \& {de la Cruz
  Rodr{\'{\i}}guez}, J. 2019, \aap, 621, A1

\bibitem[{{Rubio da Costa} \& {Kleint}(2017)}]{2017ApJ...842...82R}
{Rubio da Costa}, F. \& {Kleint}, L. 2017, \apj, 842

\bibitem[{{Rubio da Costa} {et~al.}(2015){Rubio da Costa}, {Kleint},
  {Petrosian}, {Sainz Dalda}, \& {Liu}}]{2015ApJ...804...56R}
{Rubio da Costa}, F., {Kleint}, L., {Petrosian}, V., {Sainz Dalda}, A., \&
  {Liu}, W. 2015, \apj, 804, 56

\bibitem[{{Rutten}(2017)}]{2017A&A...598A..89R}
{Rutten}, R.~J. 2017, \aap, 598, A89

\bibitem[{{Rutten} \& {Rouppe van der Voort}(2017)}]{2017A&A...597A.138R}
{Rutten}, R.~J. \& {Rouppe van der Voort}, L.~H.~M. 2017, \aap, 597, A138

\bibitem[{{Rybicki} \& {Hummer}(1991)}]{1991A&A...245..171R}
{Rybicki}, G.~B. \& {Hummer}, D.~G. 1991, \aap, 245, 171

\bibitem[{{Rybicki} \& {Hummer}(1992)}]{1992A&A...262..209R}
{Rybicki}, G.~B. \& {Hummer}, D.~G. 1992, \aap, 262, 209

\bibitem[{{Schad} {et~al.}(2013){Schad}, {Penn}, \&
  {Lin}}]{2013ApJ...768..111S}
{Schad}, T.~A., {Penn}, M.~J., \& {Lin}, H. 2013, \apj, 768, 111

\bibitem[{{Scharmer} {et~al.}(2003){Scharmer}, {Bjelksj{\"o}}, {Korhonen},
  {Lindberg}, \& {Petterson}}]{2003SPIE.4853..341S}
{Scharmer}, G.~B., {Bjelksj{\"o}}, K., {Korhonen}, T.~K., {Lindberg}, B., \&
  {Petterson}, B. 2003, in \procspie, Vol. 4853, Innovative Telescopes and
  Instrumentation for Solar Astrophysics, ed. S.~L. {Keil} \& S.~V. {Avakyan},
  International Society for Optics and Photonics, 341--350

\bibitem[{{Scharmer} {et~al.}(2008){Scharmer}, {Narayan}, {Hillberg}, {de la
  Cruz Rodriguez}, {L{\"o}fdahl}, {Kiselman}, {S{\"u}tterlin}, {van Noort}, \&
  {Lagg}}]{2008ApJ...689L..69S}
{Scharmer}, G.~B., {Narayan}, G., {Hillberg}, T., {et~al.} 2008, \apj, 689, L69

\bibitem[{{Schmit} {et~al.}(2017){Schmit}, {Sukhorukov}, {De Pontieu},
  {Leenaarts}, {Bethge}, {Winebarger}, {Auch{\`e}re}, {Bando}, {Ishikawa},
  {Kano}, {Kobayashi}, {Narukage}, \& {Trujillo Bueno}}]{2017ApJ...847..141S}
{Schmit}, D., {Sukhorukov}, A.~V., {De Pontieu}, B., {et~al.} 2017, \apj, 847,
  141

\bibitem[{{Sukhorukov} \& {Leenaarts}(2017)}]{2017A&A...597A..46S}
{Sukhorukov}, A.~V. \& {Leenaarts}, J. 2017, \aap, 597, A46

\bibitem[{{Sutton}(1978)}]{1978JQSRT..20..333S}
{Sutton}, K. 1978, \jqsrt, 20, 333

\bibitem[{{Tei} {et~al.}(2018){Tei}, {Sakaue}, {Okamoto}, {Kawate}, {Heinzel},
  {UeNo}, {Asai}, {Ichimoto}, \& {Shibata}}]{2018PASJ..tmp...61T}
{Tei}, A., {Sakaue}, T., {Okamoto}, T.~J., {et~al.} 2018, Publications of the
  Astronomical Society of Japan, 61

\bibitem[{{Uitenbroek}(2001)}]{2001ApJ...557..389U}
{Uitenbroek}, H. 2001, \apj, 557, 389

\bibitem[{{{\v S}t{\v e}p{\'a}n} {et~al.}(2015){{\v S}t{\v e}p{\'a}n},
  {Trujillo Bueno}, {Leenaarts}, \& {Carlsson}}]{2015ApJ...803...65S}
{{\v S}t{\v e}p{\'a}n}, J., {Trujillo Bueno}, J., {Leenaarts}, J., \&
  {Carlsson}, M. 2015, \apj, 803, 65

\bibitem[{{V{\"o}gler} {et~al.}(2005){V{\"o}gler}, {Shelyag}, {Sch{\"u}ssler},
  {Cattaneo}, {Emonet}, \& {Linde}}]{2005A&A...429..335V}
{V{\"o}gler}, A., {Shelyag}, S., {Sch{\"u}ssler}, M., {et~al.} 2005, \aap, 429,
  335

\end{thebibliography}



\begin{appendix} 

\section{Charge conservation in non-LTE} \label{app:charge_conservation}

\newcommand*{\rmd}{\ensuremath{\mathrm{d}}} 

In the Multi3D code, the solution of the system of statistical equilibrium
equations and the integration of the transfer equation are iterated
self-consistently using the multi-level approximate $\Lambda$-operator scheme
(M-ALI).
In the particular case described in this paper the electron density is neither
known a priori nor provided by the MURaM model atmosphere and must be inferred
along the solution of the radiative transfer problem.
For this purpose, in each M-ALI iteration we sub-iterate a system of non-linear
equations, which join statistical equilibrium and charge conservation equations
together.
So far the code can treat in kinetic equilibrium (non-LTE) only one chemical
element of interest while the others are considered in LTE.
Hydrogen is treated in non-LTE using a model atom that has three levels of
\ion{H}{I} with the principal quantum numbers $n = \{1, 2, 3\}$ and \ion{H}{II}
continuum for protons.
Here is the numerical implementation of this scheme.

The hydrogen population densities $n_i$ with $i = \{1, 2, 3\}$ and the proton
density $n_\text{p}$ satisfy the system of 3{+}1 equations of statistical
equilibrium for \ion{H}{I},
\begin{multline} \label{eq:sse_hydrogen}
  F_{\!i}
    = \dfrac{ \rmd n_i }{ \rmd t }
    \equiv
      n_\text{p}
        \bigl[ R_{\text{p},i}( n_\text{e} ) + C_{\text{p},i}( n_\text{e} ) \bigr]
      +
      \sum_{ \makebox[0pt]{$\scriptstyle{ j = 1, j \neq i }$} }^3
        n_{\!j}
          \bigl[ R_{\!j,i} + C_{\!j,i}( n_\text{e} ) \bigr]
  \\*[-4pt]
    \qquad
    - n_i
      \Bigl[
        R_{i,\text{p}}( n_\text{e} ) + C_{i,\text{p}}( n_\text{e} )
        +
        \sum_{ \makebox[0pt]{$\scriptstyle{ j = 1, j \neq i }$} }^3
          \bigl[ R_{i,j} + C_{i,j}( n_\text{e} ) \bigr]
      \Bigr]
      = 0
  \\*[-4pt]
  \quad\mbox{for }i = \{1, 2, 3\},
\end{multline}
and for \ion{H}{II},
\begin{multline} \label{eq:sse_protons}
  F_\text{\!p}
    = \dfrac{ \rmd n_\text{p} }{ \rmd t }
    \equiv
      \sum_{j = 1}^3
        n_{\!j}
          \bigl[ R_{\!j,\text{p}}( n_\text{e} ) + C_{\!j,\text{p}}( n_\text{e} ) \bigr]
  \\*[-4pt]
    - n_{\text{p}}
      \Bigl[
        \sum_{j = 1}^3
          \bigl[ R_{\text{p},j}( n_\text{e} ) + C_{\text{p},j}( n_\text{e} ) \bigr]
      \Bigr]
      = 0,
\end{multline}
where $R_{a,b}$ and $C_{\!a,b}$ correspondingly denote preconditioned radiative
and collisional rates for a transition $a {\rightarrow} b$
\citep[for details see][]{1991A&A...245..171R,1992A&A...262..209R}.
Except for the bound-bound radiative rates, $R_{i,j}$, all other rates
explicitly depend on the electron density.
We indicated this by the parentheses, $( n_\text{e} )$, and emphasized by
separating the bound-free rates having one proton subscript $\text{p}$ from the
remaining bound-bound rates having both numerical subscripts
$i, j = \{1, 2, 3\}$.

As this homogeneous system has a trivial (zero) solution, one of the equations
must be replaced with the condition of particle conservation for the total
number density of hydrogen,
\begin{equation} \label{eq:particle_conservation}
  n_\text{H}^\text{tot}
    = n_1 + n_2 + n_3 + n_\text{p}.
\end{equation}
We substitute the equation for the level having the largest population.
This makes the matrix of rates well conditioned.

If the electron density $n_\text{e}$ is known and fixed then
Eqs.\ (\ref{eq:sse_hydrogen}--\ref{eq:particle_conservation}) are linear with
respect to the unknown populations and can be solved with any appropriate
numerical method for algebraic linear systems.

If the electron density is not known then all rates explicitly dependent on
$n_\text{e}$ are unknown and the system (\ref{eq:sse_hydrogen}--%
\ref{eq:particle_conservation}) is non-linear with respect to the unknowns
$n_i$, $n_\text{p}$, and $n_\text{e}$.
We add one more equation to constrain $n_\text{e}$ by stipulating charge
neutrality, that is, the net negative and positive charges must be equal,
\begin{equation} \label{eq:charge_equality}
  n_\text{e}
    = n_\text{p} + (\mbox{charge of all positive ions but hydrogen}),
\end{equation}
which we express as
\begin{equation} \label{eq:charge_conservation}
  F_\text{\!e}
    = \dfrac{ \rmd n_\text{e} }{ \rmd t }
    \equiv
      n_\text{e}
      -
      \!\!\!\underbrace{%
        n_\text{p}%
      }_\text{non-LTE}\!\!\!
      -
      \,n_\text{H}^\text{tot}
        \underbrace{%
          \biggl[
            \sum_{s = 2}^\text{species}
              \!\!\alpha_{\!s}\!
              \sum_{i = 2}^\text{ions}
                f_{\!s,i}( n_\text{e}, T )
          \biggr]%
        }_\text{LTE}
      = 0,
\end{equation}
where for the atomic species $s$, $\alpha_{\!s}$ is the element-to-hydrogen
relative abundance, and $f_{\!s,i}$ is the corresponding ion fraction in LTE at
the ionization stage $i > 1$ relative to the neutral stage $i = 1$, which is
obtained from the Saha-Boltzmann law depending on the electron density
$n_\text{e}$ and temperature $T$.

We use the multi-dimensional Newton-Raphson method
\citep[see][Sect 9.6]{press2007numerical}
to solve this non-linear system of joined statistical equilibrium, particle, and
charge conservation equations by linearizing it to
$
  \vec{\hat J} \delta \vec{n} = -\vec{F}
$
or, in expanded form,
\begin{equation} \label{eq:newton_raphson}
  \newcommand*{\dfdnstrut}{\ensuremath{\vphantom{\frac{ \partial F_1 }{ \partial_p }}}}
  \begin{pmatrix}
    \frac{ \partial F_{\!1} }{ \partial n_1 }        &
    \frac{ \partial F_{\!1} }{ \partial n_2 }        &
    \frac{ \partial F_{\!1} }{ \partial n_3 }        &
    \frac{ \partial F_{\!1} }{ \partial n_\text{p} } &
    \frac{ \partial F_{\!1} }{ \partial n_\text{e} }
    \\
    \frac{ \partial F_{\!2} }{ \partial n_1 }        &
    \frac{ \partial F_{\!2} }{ \partial n_2 }        &
    \frac{ \partial F_{\!2} }{ \partial n_3 }        &
    \frac{ \partial F_{\!2} }{ \partial n_\text{p} } &
    \frac{ \partial F_{\!2} }{ \partial n_\text{e} }
    \\
    \frac{ \partial F_{\!3} }{ \partial n_1 }        &
    \frac{ \partial F_{\!3} }{ \partial n_2 }        &
    \frac{ \partial F_{\!3} }{ \partial n_3 }        &
    \frac{ \partial F_{\!3} }{ \partial n_\text{p} } &
    \frac{ \partial F_{\!3} }{ \partial n_\text{e} }
    \\
    \frac{ \partial F_\text{\!p} }{ \partial n_1 }        &
    \frac{ \partial F_\text{\!p} }{ \partial n_2 }        &
    \frac{ \partial F_\text{\!p} }{ \partial n_3 }        &
    \frac{ \partial F_\text{\!p} }{ \partial n_\text{p} } &
    \frac{ \partial F_\text{\!p} }{ \partial n_\text{e} }
    \\
    \frac{ \partial F_\text{\!e} }{ \partial n_1 }        &
    \frac{ \partial F_\text{\!e} }{ \partial n_2 }        &
    \frac{ \partial F_\text{\!e} }{ \partial n_3 }        &
    \frac{ \partial F_\text{\!e} }{ \partial n_\text{p} } &
    \frac{ \partial F_\text{\!e} }{ \partial n_\text{e} } %
  \end{pmatrix}
  \begin{pmatrix}
    \delta n_1        \dfdnstrut \\
    \delta n_2        \dfdnstrut \\
    \delta n_3        \dfdnstrut \\
    \delta n_\text{p} \dfdnstrut \\
    \delta n_\text{e} \dfdnstrut
  \end{pmatrix}
  =
  - 
  \begin{pmatrix}
    F_{\!1}      \dfdnstrut \\
    F_{\!2}      \dfdnstrut \\
    F_{\!3}      \dfdnstrut \\
    F_\text{\!p} \dfdnstrut \\
    F_\text{\!e} \dfdnstrut
  \end{pmatrix},
\end{equation}
where
$
  \delta\vec{n} =
  ( \delta n_1, \delta n_2, \delta n_3, \delta n_\text{p}, \delta n_\text{e} )^\top
$
is the column vector of corrections to unknowns,
$
  \vec{F} =
  ( F_\text{\!1}, F_\text{\!2}, F_\text{\!3}, F_\text{\!p}, F_\text{\!e} )^\top
$
is the column vector of free terms given by Eqs.\ \eqref{eq:sse_hydrogen},
\eqref{eq:sse_protons}, and \eqref{eq:charge_conservation}, and
$
  J_{i,j} = \partial F_{\!i}/\partial n_{\!j}
$
is the Jacobian matrix with corresponding partial derivatives.
As we already explained, one of the $F_{\!1}$, $F_{\!2}$, $F_{\!3}$, or
$F_\text{\!p}$ equations is replaced by Eq.\ \eqref{eq:particle_conservation}.
We analytically computed the partial derivatives in a fashion similar to
\citet[][see Appendix A]{2007A&A...473..625L}. 

This linear algebraic system of five equations with five unknowns is solved for
the corrections $\delta \vec{n}$ using a standard Gaussian elimination scheme
and the unknowns are updated to $\vec{n} + \delta \vec{n}$.
The initial values of $\vec{n}$ are taken from the previous M-ALI iteration and
are used to initialize $\vec{F}$ and $\vec{\hat J}$.
The process is sub-iterated until the relative change in $\max|\delta \vec{n}|$
reaches the desired tolerance of $10^{-4}$.
At each sub-iteration, only the radiative bound-bound rates $R_{i,j}$ are kept
fixed, the remaining rates and the ionization fractions, which depend on
$n_\text{e}$, are updated accordingly.
Algorithm\ \ref{alg:nested_iterations} outlines the entire scheme.
\begin{algorithm}[!t]
  \SetInd{0.5em}{0.5em}
  \DontPrintSemicolon
  \LinesNotNumbered
  \SetCommentSty{textsl}
  Initialize $\vec{n}$ using the LTE approximation.\;
  \Repeat(\tcp*[f]{M-ALI iterations}){converged $\vec{n}$}{
    Get background opacities, emisivities, and absorption profiles from
    $n_\text{e}$ as well as foreground opacities and emisivities from $n_i$ and
    $n_\text{p}$.\;
    Integrate the transfer equation for the radiation field.\;
    Re-evaluate and fix radiative bound-bound rates $R_{i,j}$.\;
    \Repeat(\tcp*[f]{non-LTE charge conservation iterations})%
           {$\max|\delta\vec{n}| / |\vec{n}| < 10^{-4}$}{
      Re-evaluate radiative and colllisional rates dependend on $n_\text{e}$:
      bound-free, $R_{i,p}(n_\text{e})$, $R_{p,i}(n_\text{e})$,
      $C_{i,p}(n_\text{e})$, $C_{p,i}(n_\text{e})$, and bound-bound
      $C_{i,j}(n_\text{e})$.\;
      Get the ionization fractions $f_{\!s,i}(n_\text{e}, T)$.\;
      Compute $\vec{F}$ and $\vec{\hat J}$ from $\vec{n}$ and the rates.\;
      Solve the system \eqref{eq:newton_raphson} for $\delta \vec{n}$.\;
      $\vec{n} \leftarrow \vec{n} + \delta \vec{n}$\;
    }
  }
  \caption{Charge conservation sub-iterations nested within M-ALI iterations.}
  \label{alg:nested_iterations}
\end{algorithm}

\end{appendix} 




\end{document}